\pdfoutput=1
\documentclass[11pt,a4paper]{article}
\usepackage[margin=25mm]{geometry}
\usepackage{amsmath,amssymb}
\usepackage{graphicx}
\usepackage{booktabs}
\usepackage{placeins}
\usepackage{caption}
\usepackage{hyperref}
\graphicspath{{figs/}}

\title{A ghost-cell immersed boundary method for the shallow water equations on Cartesian grids:
mass conservation and wetting and drying over variable topography}
\author{Yuta Yaegashi\\ \small Independent Researcher}
\date{}

\begin{document}
\maketitle

\begin{abstract}
A ghost-cell immersed boundary method is proposed for the two-dimensional shallow water equations with a
variable bed (topography), solved with
an explicit Godunov-type finite volume scheme (unsplit second-order MUSCL, HLLC flux, hydrostatic
reconstruction) on Cartesian grids. Ghost-cell methods leak mass through the wall. To find a robust and
conservative treatment, eight constructions of the ghost values and a cut-cell form of the continuity
equation were applied to seven test problems, which
include wetting and drying on sloping beds and a laboratory dam break. In the resulting scheme, the ghost
values are interpolated bilinearly at the image point; solid and dry nodes are excluded, and the weights of the
remaining nodes are renormalized. Mass is conserved by solving the continuity equation in cut-cell form, while
the momentum equations are left unchanged. The proposed scheme is about as accurate as the ghost-cell method without this modification
and preserves a lake at rest whose shoreline crosses an immersed wall. Because only the continuity equation is
modified, it avoids the treatment of the momentum equations on cut cells (wall pressure, cell merging) that a
full cut-cell method requires.
\end{abstract}

\section{Introduction}

Inundation models for tsunamis and floods must represent solid boundaries that are not aligned with the grid:
buildings and channel banks. Cartesian grids are simple and efficient, but their cell faces do not follow oblique or curved walls. Such a wall can be
replaced by a staircase of cell faces, which has neither the position nor the orientation of the wall. Immersed boundary
(IB) methods instead leave the wall where it is: the grid does not conform to it, and the wall conditions are
imposed by modifying the scheme in the cells near the wall \cite{mittal2005}. In the ghost-cell variant, this is
done by assigning values to the solid cells adjacent to the fluid \cite{tseng2003,mittal2008}. Many variants exist for compressible and incompressible flows. For the shallow
water equations (SWE), boundaries that are not aligned with the grid have been treated in several ways, which
are listed in Table~\ref{tab:prior} together with the grids and schemes on which they were used.

\begin{table}[!tp]
\centering\footnotesize
\caption{Treatments of boundaries not aligned with the grid in shallow water models (FV: finite volume; FD:
finite difference; DG: discontinuous Galerkin; FE: finite element; LBM: lattice Boltzmann method).}\label{tab:prior}
\begin{tabular}{p{0.26\linewidth}p{0.30\linewidth}p{0.36\linewidth}}
\toprule
Reference & Boundary treatment & Grid and scheme \\
\midrule
\multicolumn{3}{l}{\textbf{Ghost cells}} \\[2pt]
Liang and Borthwick \cite{liang2008} & ghost value copied from the adjacent fluid cell, velocity reflected with the local wall direction; wall placed on the cell face & adaptive quadtree, collocated, explicit Godunov-type FV \\
Liang \cite{liang2008b} & as above & uniform Cartesian, collocated, explicit Godunov-type FV \\
Rinco \cite{rinco2012} & ghost value by bilinear interpolation at the image point & uniform Cartesian, collocated, explicit central FV \\
Kwon et al.\ \cite{kwon2021} & ghost value extrapolated through the wall point from the image point; image-point value by inverse-square-distance weighting of the surrounding fluid nodes & Cartesian, collocated, explicit fifth-order WENO (FD) \\
\addlinespace[3pt]\midrule
\multicolumn{3}{l}{\textbf{Ghost cells combined with cut cells}} \\[2pt]
Canestrelli et al.\ \cite{canestrelli2016} & ghost cells for momentum, cut cells for continuity & Cartesian, staggered, implicit (ADI) FD \\
Liang et al.\ \cite{liang2007} & cut cells; ghost cells in the smallest cut cells & adaptive quadtree, collocated, explicit Godunov-type FV \\
Ning et al.\ \cite{ning2008} & as above & Cartesian, collocated, explicit Godunov-type FV (Boussinesq equations) \\
\addlinespace[3pt]\midrule
\multicolumn{3}{l}{\textbf{Cut cells}} \\[2pt]
Causon et al.\ \cite{causon2000,causon2001} & cell merging; moving walls in \cite{causon2001} & Cartesian, collocated, explicit Godunov-type FV \\
Kim and Cho \cite{kim2011} & cell merging & Cartesian, collocated, explicit Godunov-type FV \\
An and Yu \cite{an2012} & cell merging; hydrostatic reconstruction on cut cells & adaptive quadtree, collocated, explicit Godunov-type FV \\
Rosatti et al.\ \cite{rosatti2005} & cut cells in the continuity equation only; slip condition in the semi-Lagrangian trajectories & Cartesian, staggered, semi-implicit semi-Lagrangian \\
Li and Mandli \cite{li2021} & zero-width barriers, $h$-box method & Cartesian, explicit FV (wave propagation) \\
Ryoo and Mandli \cite{ryoo2021,ryoo2021b} & zero-width barriers, state redistribution or cell merging & Cartesian, explicit FV (wave propagation) \\
Taylor and Chan \cite{taylor2026} & cut meshes, state redistribution & Cartesian, high-order DG \\
\addlinespace[3pt]\midrule
\multicolumn{3}{l}{\textbf{Forcing with Lagrangian markers}} \\[2pt]
Chung et al.\ \cite{chung2011} & regularized delta function & curvilinear FD \\
Ouro et al.\ \cite{ouro2016} & direct forcing & unstructured FV \\
Oler et al.\ \cite{oler2016} & feedback and direct forcing & Cartesian (Boussinesq-type model) \\
De Rosis \cite{derosis2014}, Prestininzi et al.\ \cite{prestininzi2016} & immersed boundary forcing & LBM \\
\addlinespace[3pt]\midrule
\multicolumn{3}{l}{\textbf{Other embedded approaches}} \\[2pt]
Reckinger et al.\ \cite{reckinger2012}, Kevlahan et al.\ \cite{kevlahan2015} & Brinkman volume penalization & adaptive wavelet methods (ocean models) \\
Song et al.\ \cite{song2018} & shifted boundary method & embedded FE on a fixed background mesh \\
\bottomrule
\end{tabular}
\end{table}

Ghost-cell methods have been used on explicit, collocated finite volume and finite difference schemes
\cite{liang2008,liang2008b,rinco2012,kwon2021}. Still water was preserved in fully wet basins
\cite{liang2008,rinco2012}; mass conservation at the wall was not examined or not achieved, and wetting and
drying at the immersed
wall was not addressed. Cut-cell methods
\cite{causon2000,causon2001,kim2011} are conservative by construction. An
and Yu~\cite{an2012} implemented the hydrostatic reconstruction on cut cells with cell merging and preserved
still water in a domain in which the wet--dry interface crosses the cut cells; a
well-balanced treatment of wetting and drying at an embedded wall thus exists for cut cells. The other approaches are listed in Table~\ref{tab:prior}.

The two main families have complementary strengths. Cut-cell methods conserve mass and momentum, but they
are demanding to implement: fluxes and gradients have to be defined on cells of irregular shape, a pressure
force acts on the solid face of each cut cell, small cells have to be merged \cite{causon2000} or stabilized in
another way \cite{berger2021}, and with a variable bed the well-balanced treatment of the bed slope and of
wet--dry fronts has to be carried over to the cut cells \cite{an2012}. Ghost-cell methods avoid all
of this: the interior scheme is applied unchanged in every fluid cell, and the wall is represented only by the
values assigned to the ghost cells. Their weakness is that mass is not conserved at the wall. This paper keeps the ghost-cell method and adds to it only what is needed to conserve
mass: a cut-cell form of the continuity equation.

Hybrids of ghost cells and cut cells also exist. Liang et al.~\cite{liang2007} and Ning et
al.~\cite{ning2008} used cut cells and updated only the smallest of them with a ghost-cell method.
Cut cells have also been used for the continuity equation alone. Rosatti et al.~\cite{rosatti2005}
discretized only the continuity equation on cut cells in a semi-implicit scheme on a staggered grid and left the
pressure gradient of the momentum equations unchanged.
Canestrelli et al.~\cite{canestrelli2016}, following a method for the Navier--Stokes equations
\cite{seo2011}, combined cut cells for the continuity equation with ghost cells for the momentum equations on a
staggered grid with an implicit solver of the alternating direction type. They
reported mass conservation to machine precision, second-order spatial accuracy and no time-step limitation for
small cut cells, and found the combination more accurate than ghost cells or cut-cell continuity alone. Their
ghost values are velocities only; the water level has no ghost value on the staggered grid. All their tests
are fully wet, with a horizontal or uniformly sloping bed; wetting and drying at the immersed wall and the
preservation of still water over a variable bed are not discussed.

The scheme proposed in this paper follows Canestrelli et al.~\cite{canestrelli2016} in using ghost cells for the
momentum equations
and cut cells for the continuity equation, but differs from their scheme in three respects. First, it is explicit,
formulated for a
collocated Godunov-type scheme, whereas that of Canestrelli et al.\ is implicit. Being explicit, the present scheme is restricted by
the size of the cut cells and therefore needs a lower bound on the area fraction and a redistribution of the
remaining mass. Second, in the bilinear interpolation of the ghost values the solid nodes are simply excluded and the weights
of the remaining fluid nodes are renormalized, whereas Canestrelli et al.\ replace the value at a solid node by
the boundary condition at the nearest point of the wall; the former proved to be the more robust choice in the ghost-cell scheme without a mass treatment
(Section~\ref{sec:first}). Third, wetting and
drying at the immersed wall and the preservation of a lake at rest over a variable bed are treated, which
requires ghost values of the water level as well and the exclusion of dry nodes from them. Solid nodes
have been treated before: Rinco~\cite{rinco2012} replaced the ghost cells in the stencil by the wall conditions
and dropped the solid nodes that are not ghost cells. Dry nodes have not been excluded before. To the author's knowledge, no ghost-cell method for explicit collocated SWE solvers has been reported that
both conserves mass and preserves still water where the wet--dry front meets the immersed wall; the latter has so
far been shown only for cut cells \cite{an2012}.

In addition, the ways of constructing the ghost values have not been compared for the SWE, so that it is not
known which of them is suitable; such a comparison is therefore carried out here.

Section~\ref{sec:method} describes the scheme. Section~\ref{sec:tests} compares it with the ghost-cell scheme
without a mass treatment on seven problems, and Section~\ref{sec:summary} summarizes the comparison, including
that of the eight constructions of the ghost values. Section~\ref{sec:demo} applies the scheme to the run-up
through an array of cylinders, and
Section~\ref{sec:conclusions} gives the conclusions.

\section{Numerical method}\label{sec:method}

This section describes the proposed scheme in detail: a ghost-cell immersed boundary in which the ghost values
are interpolated bilinearly at the image point from the nodes that are neither solid nor dry, with the weights
renormalized, combined with a cut-cell form of the continuity equation,
in an unsplit second-order Godunov-type scheme. The other boundary treatments with which it is compared are
summarized in Section~\ref{sec:others}. Among the constructions of the ghost values, the renormalized bilinear
interpolation is referred to as the baseline; it is the construction used in the proposed scheme, and the
other constructions are compared against it.

\subsection{Governing equations and base scheme}\label{sec:base}

The governing equations are
\begin{equation}
\partial_t \begin{pmatrix} h \\ q_x \\ q_y \end{pmatrix}
+ \partial_x \begin{pmatrix} q_x \\ q_x^2/h + g h^2/2 \\ q_x q_y/h \end{pmatrix}
+ \partial_y \begin{pmatrix} q_y \\ q_x q_y/h \\ q_y^2/h + g h^2/2 \end{pmatrix}
= \begin{pmatrix} 0 \\[2pt]
-g h\,\partial_x z - g n_M^2\,\dfrac{q_x|\mathbf{q}|}{h^{7/3}} \\[8pt]
-g h\,\partial_y z - g n_M^2\,\dfrac{q_y|\mathbf{q}|}{h^{7/3}} \end{pmatrix},
\end{equation}
where $h$ is the depth, $\mathbf{q}=(q_x,q_y)=h\mathbf{u}$ the discharge, $z$ the bed elevation, $\eta=h+z$ the
water level, $g$ the gravitational acceleration and $n_M$ the Manning coefficient; the last terms are the
bed friction according to Manning's formula. The equations are solved on a uniform Cartesian
collocated grid.

The base scheme is an unsplit second-order finite volume scheme of MUSCL type \cite{vanleer1979} with
minmod-limited slopes, the
hydrostatic reconstruction of Audusse et al.\ \cite{audusse2004} and the HLLC flux \cite{toro1994,toro2001}.
It is well balanced: water at rest over an arbitrary bed, including a partially dry one, remains at rest.
The ghost cells are treated exactly like fluid cells in the MUSCL reconstruction and in the hydrostatic
reconstruction. Whether the scheme remains well balanced at an immersed wall therefore depends on the values
assigned to the ghost cells. The formulas of the two reconstructions, from which these values are derived, are
given in Appendix~\ref{sec:hr}.

The solution is advanced in time with the two-stage strong-stability-preserving Runge--Kutta method
\cite{shu1988,gottlieb2001},
\begin{equation}
\mathbf{U}^{(1)}=\mathbf{U}^n+\Delta t\,L(\mathbf{U}^n),\qquad \mathbf{U}^{n+1}=\tfrac12\left[\mathbf{U}^n+\mathbf{U}^{(1)}+\Delta t\,L(\mathbf{U}^{(1)})\right],
\end{equation}
where $\mathbf{U}=(h,q_x,q_y)$ and the operator $L$ contains the flux differences in both directions and the bed slope
term, all evaluated from the same state. The time step is
\begin{equation}
\Delta t=\tfrac12\,\mathrm{CFL}\,\min_{\text{wet cells}}\left(\frac{\Delta x}{|u_x|+\sqrt{gh}},\,
\frac{\Delta y}{|u_y|+\sqrt{gh}}\right),\qquad \mathrm{CFL}=0.4 .\label{eq:dt}
\end{equation}
The factor $\tfrac12$ is the stability limit of an unsplit two-dimensional scheme. The minimum is taken over all
wet cells, including the ghost cells; the area fractions of
Section~\ref{sec:massdef} do not enter. The friction term is not part of $L$; it is treated by
operator splitting: at the end of each step the discharge is multiplied by $1/(1+K\Delta t)$ with
$K=g\,n_M^2|\mathbf{q}|/h^{7/3}$.

Two thresholds are used for small depths, so that large velocities $\mathbf{q}/h$ are avoided in thin films
while as little water as possible is removed. After every Runge--Kutta stage, a depth below
$\varepsilon=10^{-12}$~m, including a negative one, is set to zero together with the discharge; this clipping
is the only operation of the scheme that is not conservative (Section~\ref{sec:mass}). In a cell with
$\varepsilon\le h<h_d$, where $h_d=10^{-6}$~m, the water is kept but
the discharge is set to zero (thin-film cut-off), and the velocity is reconstructed without slopes if any of
the three cells involved is shallower than $h_d$. The same value $h_d$ decides whether a cell is treated as dry in the
interpolation of the ghost values (Section~\ref{sec:drydef}).
\subsection{Geometry and cell classification}

The geometry is given by a signed distance function $\phi$ (negative in the fluid), with the unit normal
$\mathbf{n}=\nabla\phi/|\nabla\phi|$ pointing from the fluid into the solid. Cells with $\phi<0$ at the center are
fluid cells and are advanced in time. Ghost cells are solid-centered cells that (i) are adjacent to
a fluid cell along a grid line, (ii) lie two cells away from a fluid cell along a grid line (second layer,
needed by the reconstruction), or (iii) are cut by the wall, so that part of their area is fluid although their center lies in the solid
(called slivers below; Section~\ref{sec:massdef}). Their
values are imposed before each flux evaluation. The remaining solid cells are not used.
The wall conditions are
\begin{equation}
\mathbf{u}\cdot\mathbf{n}=0,\qquad \partial_n\left(\mathbf{u}\cdot\mathbf{t}\right)=0,\qquad \partial_n \eta=0 ,
\end{equation}
where $\mathbf{u}=(u_x,u_y)$ is the velocity vector, $\mathbf{t}$ the unit tangent to the wall and $\partial_n$ the
derivative in the direction of $\mathbf{n}$. In the scheme, the first two conditions are imposed on the
discharge $\mathbf{q}=h\mathbf{u}$ rather than on the velocity (Section~\ref{sec:bilinear}).
The water level $\eta$, rather than the depth, is mirrored; this preserves still water over a variable bed as long as
all cells are wet.

\subsection{Ghost values: renormalized bilinear interpolation}\label{sec:bilinear}

\begin{figure}[t]
\centering
\includegraphics[width=\linewidth]{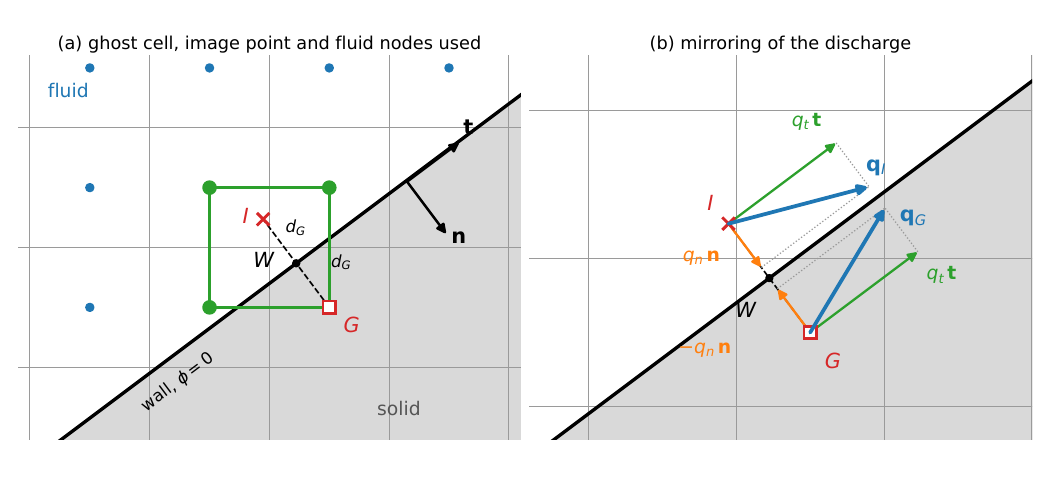}
\caption{Ghost-cell construction. (a) Ghost cell $G$, wall point $W$ and image point $I$ on the normal through
$G$. The square is the bilinear stencil of $I$, of which only the fluid cell centers (large dots) are used; here
the fourth node is $G$ itself and is dropped. (b) The discharge interpolated at $I$ is decomposed into its
normal and tangential components; at $G$ the normal component is reversed and the tangential component is
copied.}\label{fig:mirror}
\end{figure}

For a ghost cell $G$ with center $\mathbf{x}_G$ and $d_G=\phi(\mathbf{x}_G)\ge0$, the image point is
\begin{equation}
\mathbf{x}_I=\mathbf{x}_G-2\,d_G\,\mathbf{n} .
\end{equation}
Here $\mathbf{n}$ is evaluated at $\mathbf{x}_G$. If the wall is straight, $\mathbf{x}_I$ is the mirror image of
$\mathbf{x}_G$ with respect to the wall; at a curved wall it is the mirror image with respect to the tangent at
the wall point nearest to $\mathbf{x}_G$.
Let $(f_x,f_y)\in[0,1]^2$ be the position of the image point within the square formed by the four surrounding cell
centers. The bilinear weights $(1-f_x)(1-f_y)$, $f_x(1-f_y)$, $(1-f_x)f_y$ and $f_xf_y$ are non-negative and
sum to one. Because the image point of a ghost cell of the first layer lies less than one cell from the wall, one or two of the four nodes are often
ghost cells, whose values are not yet known. Only the set $F$ of fluid nodes is retained (Fig.~\ref{fig:mirror}a), and the weights are renormalized,
\begin{equation}
w_k \leftarrow \frac{w_k}{\sum_{l\in F} w_l},\qquad k\in F .\label{eq:renorm}
\end{equation}
If the sum in the denominator is below $10^{-6}$ (the image point then coincides with a non-fluid node, e.g.\ when the
wall passes through the center of the ghost cell), inverse-distance weights
$w_k\propto|\mathbf{x}_k-\mathbf{x}_I|^{-2}$ over the fluid cells within $2\Delta x$ of the image point are used
instead; the radius is enlarged until a fluid cell is found. The quantities
interpolated at the image point are the water level and the discharge,
\begin{equation}
\eta_I=\sum_k w_k\,\eta_k,\qquad \mathbf{q}_I=\sum_k w_k\,\mathbf{q}_k .\label{eq:image}
\end{equation}

\paragraph{Mirroring.}
With the values at the image point known from the interpolation, the values at the ghost cell are determined so
that the wall conditions hold at the wall point $W$ at $\mathbf{x}_W=\mathbf{x}_G-d_G\mathbf{n}$, which
is the midpoint of the segment between $G$ and $I$ (Fig.~\ref{fig:mirror}). Let
$\mathbf{n}=(n_x,n_y)$ be the unit normal at $G$ and $\mathbf{t}=(-n_y,n_x)$ the unit tangent. The
discharge at the image point is decomposed into its normal and tangential components,
\begin{equation}
\mathbf{q}_I=q_n\,\mathbf{n}+q_t\,\mathbf{t},\qquad
q_n=\mathbf{q}_I\cdot\mathbf{n}=q_{x,I}\,n_x+q_{y,I}\,n_y,\qquad
q_t=\mathbf{q}_I\cdot\mathbf{t}=-q_{x,I}\,n_y+q_{y,I}\,n_x .
\end{equation}
Each quantity is assumed to vary linearly along the normal between $G$ and $I$. Its value at the wall point $W$,
midway between them, is then the mean of the two values, and its normal derivative there is their difference
divided by $2d_G$. The three wall conditions give
\begin{align}
\text{no flow through the wall:}\quad & \tfrac12\,(q_{n,G}+q_{n,I})=0 &&\Longrightarrow\quad q_{n,G}=-q_n ,\\
\text{no shear:}\quad & \frac{q_{t,G}-q_{t,I}}{2d_G}=0 &&\Longrightarrow\quad q_{t,G}=q_t ,\\
\text{water level:}\quad & \frac{\eta_G-\eta_I}{2d_G}=0 &&\Longrightarrow\quad \eta_G=\eta_I .
\end{align}
The normal component is thus reversed, whereas the tangential component and the water level are copied. Recombining
the components and eliminating the tangential part with $q_t\mathbf{t}=\mathbf{q}_I-q_n\mathbf{n}$ gives
\begin{equation}
\mathbf{q}_G=-q_n\,\mathbf{n}+q_t\,\mathbf{t}
=\mathbf{q}_I-2\,(\mathbf{q}_I\cdot\mathbf{n})\,\mathbf{n}
=(\mathsf{I}-2\,(\mathbf{n}\otimes\mathbf{n}))\,\mathbf{q}_I ,\label{eq:reflect}
\end{equation}
where $\mathsf{I}$ is the identity matrix and $\mathbf{n}\otimes\mathbf{n}$ the dyadic product. The tangent thus need not be computed. In components,
\begin{equation}
\begin{pmatrix} q_{x,G} \\ q_{y,G} \end{pmatrix}
=\begin{pmatrix} 1-2n_x^2 & -2n_xn_y \\ -2n_xn_y & 1-2n_y^2 \end{pmatrix}
\begin{pmatrix} q_{x,I} \\ q_{y,I} \end{pmatrix}.
\end{equation}
This is the reflection of the vector $\mathbf{q}_I$ in the line parallel to the wall; it preserves the magnitude,
$|\mathbf{q}_G|=|\mathbf{q}_I|$. For a wall parallel to the $x$ axis ($\mathbf{n}=(0,\pm1)$), it reduces to
$q_{x,G}=q_{x,I}$, $q_{y,G}=-q_{y,I}$, the usual reflective boundary. Inserting (\ref{eq:image}) gives the ghost
values
\begin{equation}
\eta_G=\sum_k w_k\,\eta_k,\qquad
\mathbf{q}_G=\sum_k w_k\,(\mathsf{I}-2\,(\mathbf{n}\otimes\mathbf{n}))\,\mathbf{q}_k,\qquad h_G=\eta_G-z_G ,
\label{eq:ghost}
\end{equation}
and the velocity of the ghost cell is $\mathbf{q}_G/h_G$. The discharge, rather than the velocity, is
interpolated and mirrored; when $h_G=h_I$ (flat bed, or the bed extension of Section~\ref{sec:bed}), the velocity
is mirrored in the same way, $\mathbf{u}_G=(\mathsf{I}-2\,(\mathbf{n}\otimes\mathbf{n}))\,\mathbf{u}_I$.
The weights $w_k$ and the $2\times2$ matrices $w_k(\mathsf{I}-2\,(\mathbf{n}\otimes\mathbf{n}))$ depend only on the geometry and
are computed once; at run time, the ghost values are linear combinations of fluid values.

\paragraph{Properties.}
In all cases
\begin{equation}
w_k\ge 0,\qquad \sum_k w_k=1 ,
\end{equation}
so that the interpolated value is a convex combination of the values at the fluid nodes retained. Two properties
follow. First, the
interpolated value lies between the minimum and the maximum of these values; $\eta_G$ and the magnitude of
the discharge at the image point therefore cannot overshoot at a bore or a wet--dry front. The velocity
$\mathbf{q}_G/h_G$ is bounded as well when the water level, the bed and the discharge are formed with the same weights
(Section~\ref{sec:drydef}). Second, a uniform state is reproduced exactly. The price of the renormalization is accuracy. When a node has been dropped, the interpolated value is no longer
the value at the image point. For a linear field it is the value at the weighted mean position of the remaining
nodes, $\sum_k w_k\mathbf{x}_k$. This position is shifted from the image point toward the fluid by a distance of
order $\Delta x$, and the interpolation is therefore first-order accurate. Mirroring this value shifts the effective wall away
from the true wall in the normal direction, toward the fluid. The shift is a fraction of $\Delta x$. The wall conditions are therefore first-order accurate.

\paragraph{Number of remaining nodes.}
The method does not switch to another interpolant on the remaining nodes; the bilinear weights are merely
rescaled. With two fluid nodes on the same grid line this is one-dimensional linear interpolation along that
line, and with a single fluid node its value is copied, as for a wall on cell faces, where the method reduces to
the ordinary reflective wall. In the seven geometries of Sections~\ref{sec:tests} and~\ref{sec:demo}, four fluid
nodes are available for 55\% of the ghost cells, three for 23\%, two for 19\% and one for 2.8\%; the
inverse-distance fallback, needed only where the wall passes through the center of a ghost cell, is used for 0.7\%.

\subsection{Bed elevation in the ghost cells}\label{sec:bed}

The depth of a ghost cell is $h_G=\eta_G-z_G$, so a bed elevation $z_G$ is needed inside the wall, where it
has no physical meaning. The bed is extended from the fluid side with the weights used for the water level,
\begin{equation}
z_G=\sum_k w_k z_k \quad\Longrightarrow\quad
h_G=\eta_G-z_G=\sum_k w_k\left(\eta_k-z_k\right)=\sum_k w_k h_k ,\label{eq:bedext}
\end{equation}
so that not only the water level but also the depth of the ghost cell equals the interpolated value at the image
point, and the velocity of the ghost cell is the mirrored velocity at the image point. This static extension is used when all
nodes are wet; Section~\ref{sec:drydef} gives the general case. The bed elevations stored in the
terrain data inside the walls are therefore not used, and the result does not depend on them.

If the value stored in the terrain data were used as $z_G$ instead, the water level of the ghost cell would still
be the interpolated one, but its depth would differ from that at the image point, and so would its velocity
$\mathbf{q}_G/h_G$: a bed higher than at the fluid nodes gives a shallower ghost cell and a larger velocity.

\subsection{Dry cells}\label{sec:drydef}

In a dry cell $h=0$, so the stored value $\eta=z$ is the bed elevation, not a water level. On a sloping bed it
lies above the level of the neighboring water. If such a cell enters the interpolation of $\eta$, the water
level of the ghost cell is raised above that level; the ghost cell then holds water that does not exist, and a
lake at rest whose shoreline crosses the wall is set in motion. The ghost values are therefore
formed from the wet nodes only. This is the renormalization of Section~\ref{sec:bilinear} applied to the dry nodes.
The difference is that the solid nodes are fixed, whereas the wet nodes change in time, so that the weights are
renormalized anew in every Runge--Kutta stage. Let $\mathcal{W}$ be the set of the fluid nodes used for the ghost cell that have $h_k\ge h_d$ at
the current stage, and $S=\sum_{k\in\mathcal{W}}w_k$ the sum of their weights. If $S\le0.05$, the wet nodes carry almost no weight, which is the case when the image
point lies on the dry side of the shoreline; the ghost cell is then dry (see below). Otherwise,
\begin{equation}
\eta_G=\frac1S\sum_{k\in\mathcal{W}}w_k\eta_k,\qquad
z_G=\frac1S\sum_{k\in\mathcal{W}}w_kz_k,\qquad
\mathbf{q}_G=\frac1S\sum_{k\in\mathcal{W}}w_k\,(\mathsf{I}-2\,(\mathbf{n}\otimes\mathbf{n}))\,\mathbf{q}_k .\label{eq:dryghost}
\end{equation}
The same weights are thus used for the water level, the bed and the discharge. (In the code the sum for the
discharge runs over all nodes, which gives the same value because the discharge of the excluded nodes is zero;
Section~\ref{sec:base}.) It follows
that the depth of the ghost cell is the weighted mean of the wet depths,
$h_G=\eta_G-z_G=S^{-1}\sum_{\mathcal{W}}w_kh_k\ge h_d$, and that its velocity $\mathbf{q}_G/h_G$ is the reflection of
the depth-weighted mean of the velocities of the wet nodes, which cannot exceed the largest of them in
magnitude. If the water level alone were renormalized and the bed and the discharge were kept as in
Sections~\ref{sec:bilinear} and~\ref{sec:bed}, $h_G$ could become much smaller than the depths from which $\mathbf{q}_G$ is formed, and
the velocity of the ghost cell would be unbounded; this was observed to cause divergence at fronts on sloping
beds. With (\ref{eq:dryghost}) the bed elevation of a ghost cell depends on which of its nodes are wet and
therefore changes in time at a moving shoreline.

A ghost cell is dry if $S\le0.05$; in addition, a ghost cell of the first layer is dry if
none of the fluid cells that share a face with it has $h\ge h_d$ (a cell of the second layer shares no face with
a fluid cell). A dry ghost cell has $h_G=0$ and $\mathbf{q}_G=0$; what remains to be chosen is its
bed elevation. In the first case it is the static value $z_G^{\mathrm{s}}$ of Section~\ref{sec:bed},
because the few wet nodes do not determine a water level to which the bed could be raised. In the
second case it is raised to the
interpolated water level, $z_G=\max(z_G^{\mathrm{s}},\eta_G)$ with $\eta_G$ from (\ref{eq:dryghost}). Without this, a
ghost cell that has been made dry can lie below the water level of a neighboring wet ghost cell; water then
flows into it, is recorded in its book value and, at the end of the step, is transferred to the neighboring
fluid cells (Section~\ref{sec:massdef}), which are dry; a
lake at rest is thereby set in motion for some positions of the wall (Section~\ref{sec:lake}). The same could in
principle happen in the first case, since $z_G^{\mathrm{s}}$ is a mean over all nodes and can lie slightly below
the water level when up to 5\% of the weight is on wet nodes; this did not occur in the tests, but it has not been
excluded. The treatment is applied to both layers of ghost cells and in both Runge--Kutta
stages.

\subsection{Cut-cell continuity equation}\label{sec:massdef}

\begin{figure}[t]
\centering
\includegraphics[width=\linewidth]{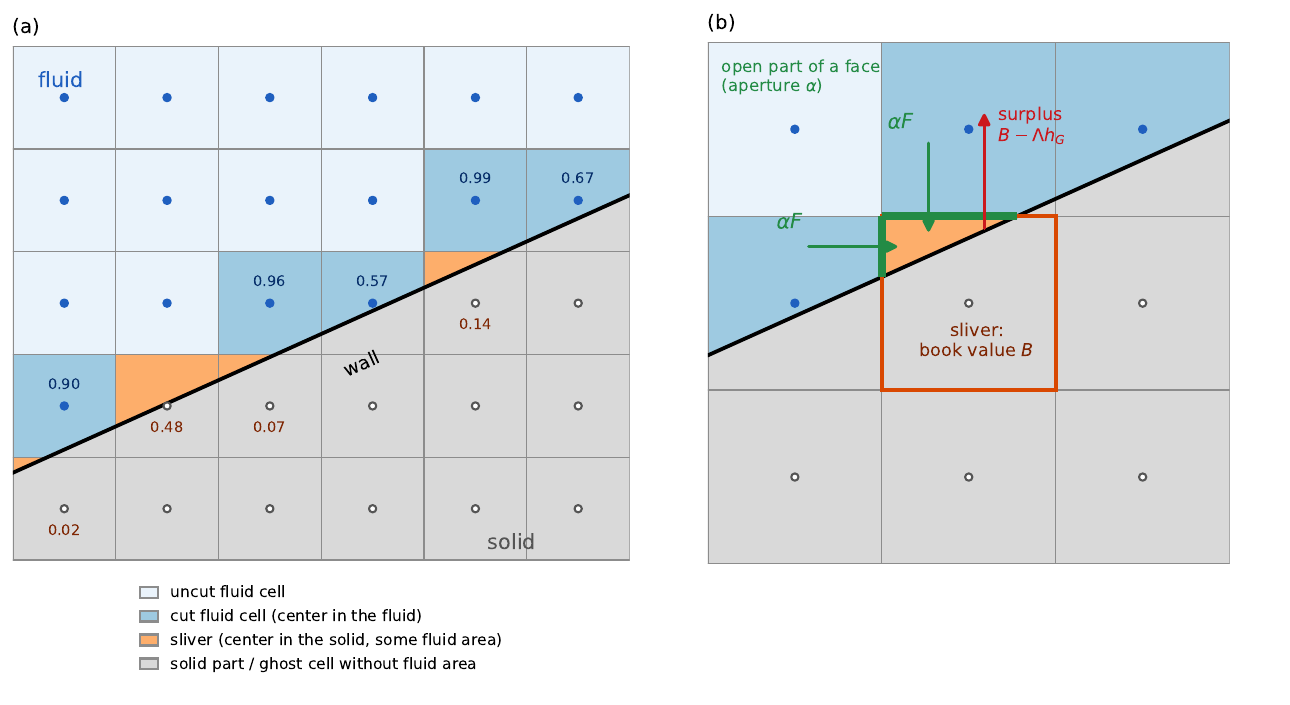}
\caption{Cells cut by the wall. (a) Classification; the numbers are the area fractions $\Lambda$, and the dots are
the cell centers (filled: in the fluid; open: in the solid). (b) One sliver (orange frame): its book value $B$
changes by the mass fluxes through the open parts of its faces (green; aperture $\alpha$), and at the end of the
step the surplus $B-\Lambda h_G$ is transferred to the neighboring fluid cells.}\label{fig:sliver}
\end{figure}

In a ghost-cell method, the numerical mass flux through the faces between fluid and ghost cells is not zero, and
the cells next to the wall are treated as full cells although part of each is solid. Both effects cause the
volume to change in time. They are removed by solving only the continuity equation in cut-cell form, while the
velocity is advanced by the ghost-cell scheme. The cut-cell form uses two geometric quantities, which are computed
once from $\phi$: the area fraction $\Lambda$ of each cell (the part of its area that is fluid) and the aperture
$\alpha$ of each face (the part of its length that is open). Both are obtained by locating the wall by
bisection, on the faces and on 64 lines across each cell. A fluid cell with $\Lambda<1$ or with a face of
$\alpha<1$ is called a cut fluid cell (Fig.~\ref{fig:sliver}a).

\paragraph{Cut fluid cells.}
For a cut fluid cell, the computation proceeds in three steps.
\begin{enumerate}
\item Ghost-cell step. The ghost values are imposed, and the mass and momentum fluxes at the four faces are
computed. The time derivatives are then formed as in an uncut cell, i.e.\ with the full faces and the full cell
area; they are denoted by $(\partial_t h)^{\mathrm{gc}}$ and $(\partial_t\mathbf{q})^{\mathrm{gc}}$.
\item Cut-cell step, depth only. The same mass fluxes $F$ are reused; no new flux is computed. They are
weighted by the apertures and divided by the area fraction, which gives the time derivative denoted by cc,
\begin{equation}
(\partial_t h)^{\mathrm{cc}}=\frac{M}{\max(\Lambda,\Lambda_{\min})},\qquad
M=-\frac{\alpha_eF_e-\alpha_wF_w}{\Delta x}-\frac{\alpha_nF_n-\alpha_sF_s}{\Delta y} .\label{eq:hcc}
\end{equation}
Here $F_e$, $F_w$, $F_n$ and $F_s$ are the mass fluxes through the east, west, north and south faces of the cell,
and $\alpha_e$, $\alpha_w$, $\alpha_n$ and $\alpha_s$ are the apertures of these faces.
\item Replacement. The time derivative of the depth is replaced by $(\partial_t h)^{\mathrm{cc}}$. That of the
discharge is changed so that the time derivative of the velocity remains that of the ghost-cell scheme,
\begin{equation}
(\partial_t\mathbf{u})^{\mathrm{cc}}=(\partial_t\mathbf{u})^{\mathrm{gc}} .\label{eq:ureq}
\end{equation}
\end{enumerate}
Since $\mathbf{q}=h\mathbf{u}$,
$\partial_t\mathbf{q}=\mathbf{u}\,\partial_t h+h\,\partial_t\mathbf{u}$, and with (\ref{eq:ureq})
\begin{equation}
(\partial_t\mathbf{q})^{\mathrm{cc}}=\mathbf{u}\,(\partial_t h)^{\mathrm{cc}}
+h\,(\partial_t\mathbf{u})^{\mathrm{gc}} ,\label{eq:qsplit}
\end{equation}
where the first term comes from the cut-cell form and the second from the ghost-cell scheme. Eliminating the
second term with
$h\,(\partial_t\mathbf{u})^{\mathrm{gc}}=(\partial_t\mathbf{q})^{\mathrm{gc}}-\mathbf{u}\,(\partial_t h)^{\mathrm{gc}}$
gives the form used in the computation,
\begin{equation}
(\partial_t\mathbf{q})^{\mathrm{cc}}=(\partial_t\mathbf{q})^{\mathrm{gc}}
+\mathbf{u}\left[(\partial_t h)^{\mathrm{cc}}-(\partial_t h)^{\mathrm{gc}}\right].\label{eq:qcorr}
\end{equation}
Without this correction, the new velocity $\mathbf{q}/h$ would change merely because $h$ was updated
differently.

\paragraph{Lower bound of the area fraction.}
The bound $\Lambda_{\min}$ in (\ref{eq:hcc}) avoids the small-cell problem: for a given mass flux, the depth
of a cell with area fraction $\Lambda$ changes $1/\Lambda$ times as much as that of a full cell. The mass that does not enter the cell because of the bound, a fraction $1-\Lambda/\Lambda_{\min}$ of $M\Delta t$, is kept in a pool belonging
to the cell and redistributed at the end of the step (see below), so that the scheme remains conservative.
The bound and the redistribution of the pool together correspond to the flux redistribution of cut-cell
methods~\cite{chern1987,pember1995}, applied here to the continuity equation alone.

The bound applies to the fluid cells only. A sliver, whose area fraction can be arbitrarily small, needs no
bound: its depth is not obtained by dividing an amount of water by its fluid area but is the interpolated ghost
value $h_G$, and the water that enters or leaves it is accumulated as an amount, without a division by the area
(see below). A fluid cell has its center in the fluid, so that along a straight wall its area fraction lies between 0.5 and
1; smaller values occur only in the few cells in which the wall has a corner. The bound is therefore chosen
between 0.5 and 1, and $\Lambda_{\min}=0.75$ is used for all results of this paper. With this value, the depth of a cut cell changes at most $4/3$ times as much as that of a full cell, and along a
straight wall at most one third of the mass flowing into a cell is redistributed. The results are not sensitive to this value: when the seven problems of Section~\ref{sec:tests} were
repeated with $\Lambda_{\min}=0.5$, all completed, and the errors changed by less than 1\%.

\paragraph{Slivers.}
The amount of water held by a solid-centered cell that contains fluid area (a sliver; Fig.~\ref{fig:sliver}b) is
counted in two
ways. The book value $B$ is the amount held initially, $\Lambda h_G$, plus the water that has entered or
left through the cell faces, accumulated step by step with the weights of the Runge--Kutta stages. The flux
through a face is the aperture times the HLLC mass flux that was computed at that face from the imposed ghost
states; no additional flux is evaluated. Faces
between two slivers are included, so that water can pass along the wall from sliver to sliver. A book value is
kept for every ghost cell; for a ghost cell without fluid area, $\Lambda=0$ and all apertures vanish, so that it
remains zero. The
interpolated value is the area fraction times the interpolated depth, $\Lambda h_G$, i.e.\ the amount
that corresponds to the depth assigned to the sliver by the ghost-cell method. At the end of each step, the
difference $B-\Lambda h_G$ (the surplus, which may be negative) is transferred to the neighboring fluid cells,
so that the book value again equals the interpolated value; $h_G$ is the value imposed at the beginning of the
second stage. The surplus is not bounded by any property of the scheme, because $h_G$ is an interpolated value
and not the solution of a conservation law.

The transfer is necessary because nothing else returns this water to the flow. In a fluid cell, water that
accumulates raises the depth and is driven out by the pressure. In a sliver, the fluxes are computed from the
interpolated depth $h_G$, on which the book value has no influence. Without the transfer, the water that passes
through the faces between the fluid cells and the slivers would therefore accumulate in the book values: it would
be counted, but lost to the flow. This is the leak of the ghost-cell method, which the transfer removes.

\paragraph{Redistribution.}
Redistribution takes place once per step, after the second stage and after the friction step. The surpluses of
the slivers are redistributed first and the pools second. An amount of water $A$ from a sliver or a pool,
expressed as its volume divided by the cell area $\Delta x\,\Delta y$, is passed to the neighboring cells by
the following rules.
\begin{itemize}
\item \emph{Receiving cells.} $A$ is distributed among the fluid cells that share a face with the cell, in
proportion to their water volume $\Lambda h$ (plus $10^{-12}$~m when $A>0$, so that dry neighbors can receive).
For a pool, only neighbors with $\Lambda\ge\Lambda_{\min}$ receive.
\item \emph{Depth.} A receiving cell changes its depth by its share divided by $\max(\Lambda,\Lambda_{\min})$
and adds the remaining fraction $1-\Lambda/\max(\Lambda,\Lambda_{\min})$ of the share to its own pool.
\item \emph{Velocity.} The velocity of the receiving cell is kept, so that momentum is exchanged without a
momentum flux. This rule was chosen because it is simple and robust.
\item \emph{Negative amounts.} The amount $A$ is negative when more water has left a sliver than has entered,
or when the bound $\Lambda_{\min}$ has kept the depth of a cut fluid cell from falling as far as the outflow
requires. The missing water is then taken from the receiving cells by the rules above. So that no cell is
emptied, at most half of their water volume is taken in one step; the rest remains in the book value of the
sliver, or in the pool, and is taken in a later step, when the neighbors hold more water.
\item \emph{Slivers without a fluid neighbor.} A sliver none of whose four neighbors is a fluid cell passes its
surplus in equal parts to the book values of the neighboring ghost cells, from which it is redistributed in turn.
\end{itemize}
The cells are visited in a fixed order. Because each transfer changes the depths seen by the next one, the
result depends on this order; the amounts transferred are given in Section~\ref{sec:mass}. The redistribution is
applied outside the Runge--Kutta stages and is first order in time. The thin-film cut-off is not applied again
after the redistribution, so that a cell can end the step with $\varepsilon\le h<h_d$ and a non-zero discharge;
the bound on the velocity of the ghost cells stated in Section~\ref{sec:drydef} then holds only approximately in
the next stage.

\paragraph{Conserved volume.}
With these rules the quantity
\begin{equation}
V=\Delta x\,\Delta y\left[\sum_{\text{fluid cells}}(\Lambda h+\text{pool})+\sum_{\text{slivers}}B\right]
\end{equation}
changes only through the open boundaries, apart from the clipping of
Section~\ref{sec:base}.

\paragraph{Unsplit update.}
The aperture form requires the fluxes through all faces of a cell to be differenced from the same state:
for a uniform flow along an oblique wall, the $x$ and $y$ contributions to $M$ do not vanish separately; only
their sum does. The update is therefore unsplit; with dimensional splitting, this form of the continuity
equation was found to be unstable.

\subsection{Algorithm}

One time step consists of the following operations; steps 1--5 are carried out in each of the two
Runge--Kutta stages.
\begin{enumerate}
\item Impose the ghost values (\ref{eq:ghost}), with the dry-cell treatment (\ref{eq:dryghost}).
\item Reconstruct, apply the hydrostatic reconstruction and compute the HLLC fluxes at all faces.
\item Form the time derivatives of all fluid cells; in cut fluid cells, replace them by (\ref{eq:hcc}) and
(\ref{eq:qcorr}), with $\mathbf{u}$ of the current stage, and add the part of $M\Delta t$ that does not enter the cell to the pool
with the weight $\tfrac12$ of the stage.
\item Add the aperture-weighted mass fluxes to the book values of the ghost cells, with the same weight.
\item Update the fluid cells; in the second stage, form the average with the old state and apply the friction
factor. Apply the two thresholds of Section~\ref{sec:base} and recompute $\mathbf{u}=\mathbf{q}/h$.
\item After the second stage: return the surplus of the slivers, then redistribute the pools.
\end{enumerate}
The additional work relative to the plain ghost-cell scheme is confined to the cells along the wall and
involves no system of equations. In the present code the cut-cell continuity equation with its redistribution
and the computation of the fractions take about 200 lines, compared with about 300 for the ghost-cell part.
The time per cell and step was about 7--8\% larger than without the cut-cell continuity equation in the $45^\circ$ bend
($494\times168$ cells) and in the vortex ($\Delta x=0.05$~m, $240\times240$ cells; eight threads, smallest of three runs on an
otherwise idle machine). The three runs scatter by 2\% in the bend and by up to 14\% in the vortex, so that
only the order of magnitude of the overhead is established. Ghost cells make up 4--7\%,
cut fluid cells 1--4\% and slivers about 1\% of the number of fluid cells in these two problems.

\subsection{Other boundary treatments considered}\label{sec:others}

The baseline is compared with seven other constructions of the ghost values
(Table~\ref{tab:methods}). All of them impose values in ghost cells on the solid side and fit the same
framework: the ghost values are linear combinations
$\eta_G=\sum_k w_k\eta_k$, $\mathbf{q}_G=\sum_k\mathsf{W}_k\mathbf{q}_k$ with weights and $2\times2$ matrices
$\mathsf{W}_k$ that depend only on the geometry, so that all variants share the same code at run time and
differ only in how the weights are built.

Reconstructions on the fluid side
\cite{fadlun2000} and extrapolation along grid lines, which cannot be combined with the second-order scheme used
here, are not considered. The variants differ in four properties, which are listed in
Table~\ref{tab:methods}: whether an image point is used, whether the weights are non-negative, whether linear
fields are reproduced (formal order at the wall), and whether a linear system has to be solved for each ghost cell
during set-up. The sign of the weights proves decisive for robustness (Section~\ref{sec:first}). The bed extension of
Section~\ref{sec:bed} is applied to the variants that use an image point; the WLS variants
use the bed elevation as given.

Interpolating the velocity instead of the discharge in the baseline was also tested. Within the proposed scheme, the errors
changed by less than 3\% in the seven problems. Without the cut-cell continuity equation, the leak increased in two of the problems with a dry bed (from 3.2 to 5.2\% in the $45^\circ$ bend and from 34 to 39\% in the network),
and the largest difference at a gauge in the bend increased by 22\%. This variant is not considered further.

\begin{table}[t]
\centering\small
\caption{Boundary treatments for the ghost values. Image point: whether the value is interpolated at a point on
the fluid side and reflected. Weights: $+$ non-negative, $\pm$ can be negative.
Linear: whether linear fields are reproduced (partly: only if no node is dropped). System: whether a linear system is solved per ghost cell at set-up.
The constructions are described in Appendix~\ref{sec:otherdef}.}\label{tab:methods}
\setlength{\tabcolsep}{3pt}
\begin{tabular}{p{0.19\linewidth}p{0.40\linewidth}cccc}
\toprule
Name & Construction of the boundary value & Image point & Weights & Linear & System \\
\midrule
G-bilinear (baseline; used in the proposed scheme) & bilinear interpolation at the image point; non-fluid nodes dropped, weights
renormalized (Section~\ref{sec:bilinear}) & yes & $+$ & partly & no \\
\addlinespace
G-bilinear, fixed probe & as above, but the interpolation point is at $1.5\Delta x$ from the wall & yes & $+$ & partly & no \\
\addlinespace
G-constrained bilinear & bilinear interpolation at the image point; non-fluid nodes replaced by the wall
conditions at their projections on the wall \cite{mittal2008,rinco2012} & yes & $\pm$ & yes & yes \\
\addlinespace
G-IDW & inverse-distance weighting \cite{shepard1968,gao2007} of the fluid cells within $2\Delta x$ of the image
point & yes & $+$ & no & no \\
\addlinespace
G-MLS1, G-MLS2 & moving least squares \cite{lancaster1981,vanella2009} of degree 1 or 2 over the fluid cells
within $2.5\Delta x$ of the image point & yes & $\pm$ & yes & yes \\
\addlinespace
G-WLS1, G-WLS2 & weighted least squares in wall coordinates with a basis that satisfies the wall conditions,
evaluated at the ghost cell & no & $\pm$ & yes & yes \\
\bottomrule
\end{tabular}
\end{table}

\section{Test problems and results}\label{sec:tests}

\FloatBarrier
\subsection{Problems and schemes}

The test problems are listed in Table~\ref{tab:tests}. Each problem has its own subsection, which gives the set-up
(geometry, initial condition, boundary conditions and grid spacing), the reference, and the results of the
compared schemes. Open boundaries lie at the edge of the domain. At an inflow either the full state is prescribed (oblique jump)
or a uniform unit discharge, with the depth taken from the interior but not less than the critical depth
(network); outflows have a zero gradient or a free overfall. All other boundaries are walls;
a wall is immersed unless it is stated to lie on a grid line. A wall on a grid line at the edge of the domain
is imposed with two layers of mirror cells outside it (depth and bed elevation copied, normal velocity reversed),
so that the two reconstructed states at the wall face are mirror images of each other and no water crosses the
wall. The gravitational acceleration is
$g=9.8$~m/s$^2$, the cells are square, and the initial velocity is zero unless stated otherwise. The signed
distance $\phi$ is evaluated analytically in the channel problem and is interpolated bilinearly from
its values at the cell centers in the others. In all problems the center of the geometry is offset from the
grid lines by a fraction of a cell, so that the grid is not symmetric about it.

The channel problem is computed on a dry and on a wet bed and is counted twice, which gives
seven problems. Two schemes are compared in every problem; both use the unsplit second-order scheme of
Section~\ref{sec:base} and the renormalized bilinear interpolation of
Section~\ref{sec:bilinear}.
\begin{enumerate}
\item Ghost cell: the ghost values of Section~\ref{sec:bilinear} are imposed, with neither a treatment of the
wall mass flux nor the dry-cell treatment. This is called the ghost-cell scheme below.
\item Cut-cell continuity: the ghost-cell scheme with the dry-cell treatment of Section~\ref{sec:drydef} and
the cut-cell continuity equation (Section~\ref{sec:massdef}). This is the proposed scheme of
Section~\ref{sec:method}.
\end{enumerate}
The two schemes above use the baseline interpolation. In addition, the eight ways of constructing the ghost
values (Table~\ref{tab:methods}) are compared in both schemes. For the ghost-cell scheme, the range of the
results and the number of diverged runs are given for each problem; the results for the proposed scheme are
summarized in Section~\ref{sec:summary}. The effect of the dry-cell treatment alone, without the cut-cell
continuity equation, is examined in the lake at rest (Section~\ref{sec:lake}) and in Section~\ref{sec:dry}.

Within each problem, all schemes and variants were run under identical conditions: the same grid, initial and
boundary conditions, CFL number (0.4), limiter and
thresholds. Only the treatment of the immersed wall
differs. No parameter was adjusted for an individual scheme. A run is counted as diverged if the depth
or the discharge becomes undefined or if the time step falls below one thousandth of its initial value.

\begin{table}[!htb]
\centering\small
\caption{Test problems.}\label{tab:tests}
\begin{tabular}{p{0.15\linewidth}p{0.37\linewidth}ccp{0.21\linewidth}}
\toprule
Problem & Description & $\Delta x$ [m] & $t_{\mathrm{end}}$ [s] & Reference \\
\midrule
Lake at rest & basin rotated by $30^\circ$ with a bed sloping in two directions; the shoreline crosses the walls & 0.05 & 20 & rest \\
Channel (dry, wet) & dam break in a $20\times2$~m channel rotated by $45^\circ$ (0--75$^\circ$ in Fig.~\ref{fig:angle}) & 0.093 & 1.5 & exact \\

Oblique jump & $Fr=2.74$ stream deflected by $8.95^\circ$ & 0.2 & 15 & theory ($30.01^\circ$, 1.4999~m) \\
Vortex & rigid rotation in a circular basin, wall speed 2~m/s, two revolutions & 0.1 & 31.4 & exact steady state \\
Network & flood wave entering a dry looped channel network on a slope & 0.0625 & 60 & --- (robustness, volume) \\
$45^\circ$ bend & dam break in a channel with a $45^\circ$ bend, dry bed (CADAM) & 0.02 & 40 & measurements; \cite{lai2011} \\
\bottomrule
\end{tabular}
\end{table}

\FloatBarrier
\subsection{Lake at rest}\label{sec:lake}

\begin{figure}[!htb]
\centering
\includegraphics[width=0.44\linewidth]{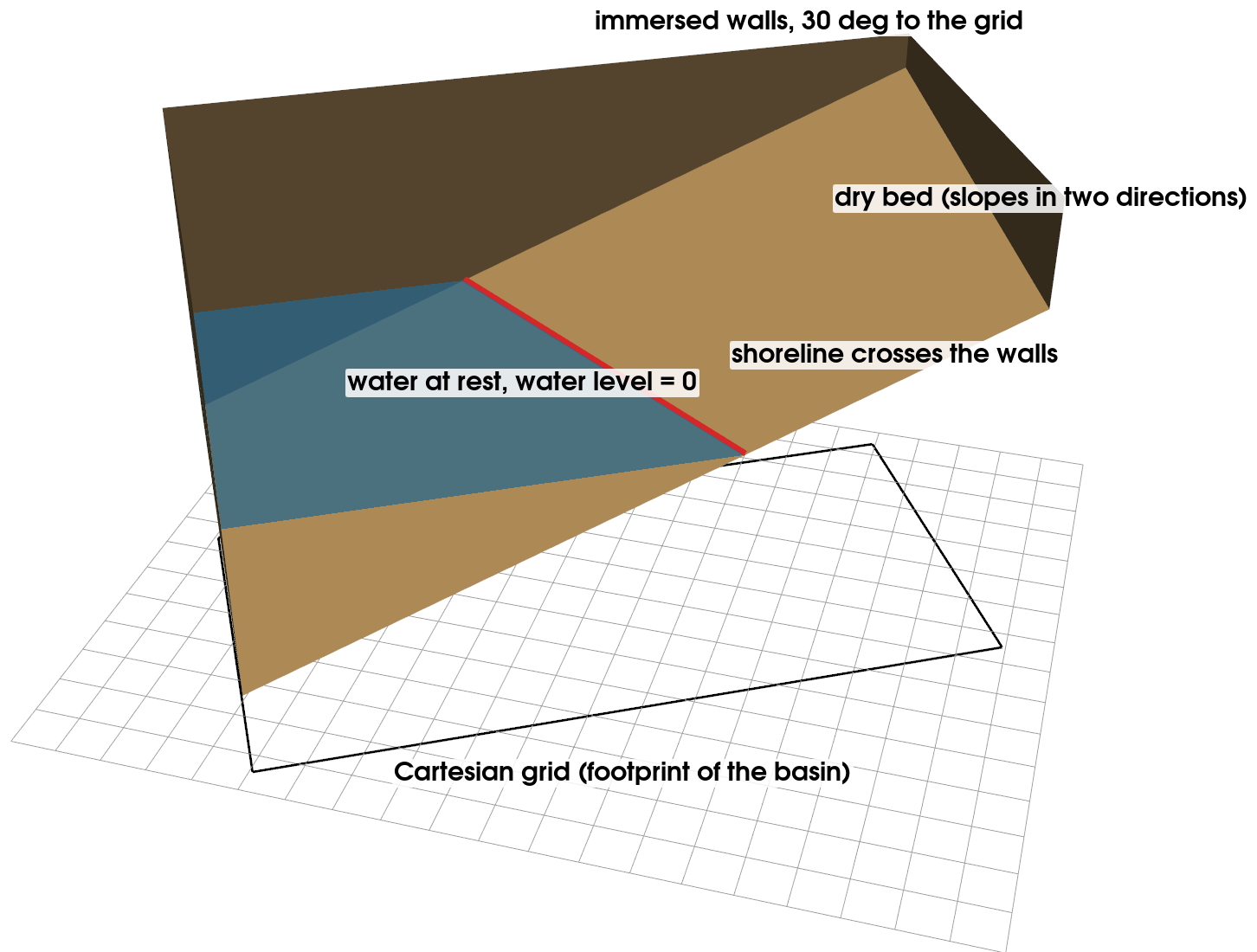}\hspace{0.04\linewidth}%
\includegraphics[width=0.40\linewidth]{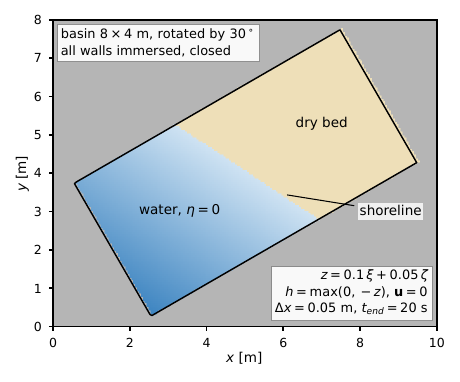}
\caption{Lake at rest. Left: perspective view (vertical scale exaggerated by a factor of 4; the two walls in front
are omitted; red: shoreline). Right: set-up (gray: solid; blue: water at $t=0$, darker where deeper; sand: dry
bed).}\label{fig:plake}
\end{figure}

The bed of a basin rotated by $30^\circ$ slopes in two directions, so the shoreline of the water at
rest crosses the immersed walls (Fig.~\ref{fig:plake}). The basin measures $8\times4$~m; with $\xi$ and $\zeta$
along its long and short axes, measured from its center, the bed is $z=0.1\,\xi+0.05\,\zeta$ and the water level is
$\eta=0$, so that the shoreline meets the two long walls. The exact solution is the state of rest, and the error
is the maximum speed after 20~s in the cells deeper than $10^{-4}$~m.

\begin{table}[!htb]
\centering\small
\caption{Lake at rest: volume balance error and maximum speed after 20~s.}\label{tab:plake}
\begin{tabular}{lcc}
\toprule
Scheme & balance [\%] & max.\ speed [m/s] \\
\midrule
Ghost cell & $1.1\times10^{-3}$ & 0.0504 \\
Ghost cell + dry-cell treatment & 0 & $1.9\times10^{-14}$ \\
Cut-cell continuity (proposed) & $2.7\times10^{-14}$ & $1.8\times10^{-14}$ \\
\midrule
Ghost cell, eight constructions (0 diverged) & $7.9\times10^{-4}$--0.0102 & 0.0249--1.06 \\
\bottomrule
\end{tabular}
\end{table}

The ghost-cell scheme generates spurious velocities of 0.05~m/s because dry cells, whose
water level is that of the bed, enter the interpolation of the water level (Table~\ref{tab:plake}). Among the eight constructions, the spurious speed ranges from 0.02 to 1.1~m/s. With the dry-cell
treatment, the speed stays at $10^{-14}$~m/s, with or without the cut-cell continuity equation (second and
third rows of Table~\ref{tab:plake}).

Since the result may depend on where the shoreline meets the wall, the computation with the proposed scheme was
repeated with the basin rotated by every degree from 1 to $89^\circ$ (89 runs of 1~s), with the water level
changed by $\pm0.05$ and $\pm0.15$~m, and with the basin shifted by four fractions of a cell (20~s each). The
largest speed in all of these runs is $3\times10^{-14}$~m/s. This property is not trivial: without the raising of
the bed of the ghost cells that are made dry (Section~\ref{sec:drydef}), the lake was set in motion at six of
the 89 angles (55, 56, 64, 65, 71 and $75^\circ$), with speeds of up to 0.76~m/s in a film thinner than 1~mm and
$2.5\times10^{-3}$~m/s where the depth exceeds 1~mm, although it stayed at rest at $30^\circ$. The preservation
of the lake at rest is therefore an observation over the configurations tested and not a proven property of the
scheme.

\FloatBarrier
\subsection{Channel: dam break in an oblique channel}\label{sec:channel}

\begin{figure}[!htb]
\centering
\includegraphics[width=0.74\linewidth]{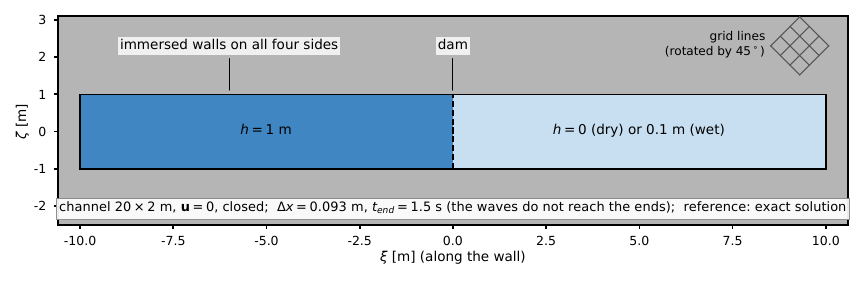}\\[2pt]
\includegraphics[width=0.42\linewidth]{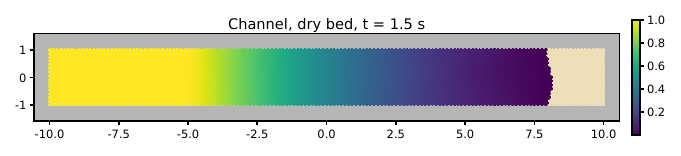}\hfill
\includegraphics[width=0.42\linewidth]{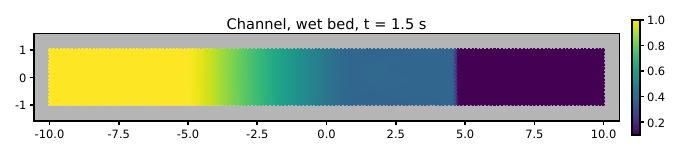}
\caption{Channel. Top: set-up in coordinates aligned with the wall (gray: solid; blue: water at $t=0$). Bottom: water level $\eta$ [m] at $t=1.5$~s computed with the proposed scheme, on a dry and on a wet bed.}\label{fig:pchannel}
\end{figure}

A channel of $20\times2$~m is rotated by $45^\circ$ with respect to the grid, and a dam at its midpoint separates
a depth of 1~m from a dry bed or from a depth of 0.1~m (Fig.~\ref{fig:pchannel}). The flow is one-dimensional
along the channel, so the exact solution of the dam-break problem \cite{toro2001} serves as the reference. The error is the mean
absolute error of the depth at $t=1.5$~s.

\begin{table}[!htb]
\centering\footnotesize
\caption{Channel: volume balance error and mean errors of the depth, the along-channel velocity and the
along-channel discharge at $t=1.5$~s ($\Delta x=0.093$~m). The velocity error is taken over the cells deeper than
1~mm.}\label{tab:pchannel}
\setlength{\tabcolsep}{4pt}
\begin{tabular}{lcccc}
\toprule
Scheme & balance & depth & velocity & discharge \\
 & [\%] & [$10^{-3}$~m] & [m/s] & [$10^{-3}$~m$^2$/s] \\
\midrule
\multicolumn{5}{l}{Dry bed} \\
\quad Ghost cell & 0.0222 & 2.99 & 0.0178 & 8.09 \\
\quad Cut-cell continuity (proposed) & $3.4\times10^{-11}$ & 3.04 & 0.0179 & 8.28 \\
\quad Ghost cell, eight constructions (0 diverged) & 0.0116--0.0334 & 2.96--3.04 & 0.0177--0.0183 & 7.63--8.27 \\
\midrule
\multicolumn{5}{l}{Wet bed} \\
\quad Ghost cell & 0.0384 & 2.96 & 0.0152 & 6.06 \\
\quad Cut-cell continuity (proposed) & $4.2\times10^{-12}$ & 3.05 & 0.0156 & 6.16 \\
\quad Ghost cell, eight constructions (0 diverged) & 0.0193--0.0675 & 2.88--2.99 & 0.0150--0.0154 & 5.74--6.38 \\
\bottomrule
\end{tabular}
\end{table}

\begin{figure}[!htb]
\centering
\includegraphics[width=\linewidth]{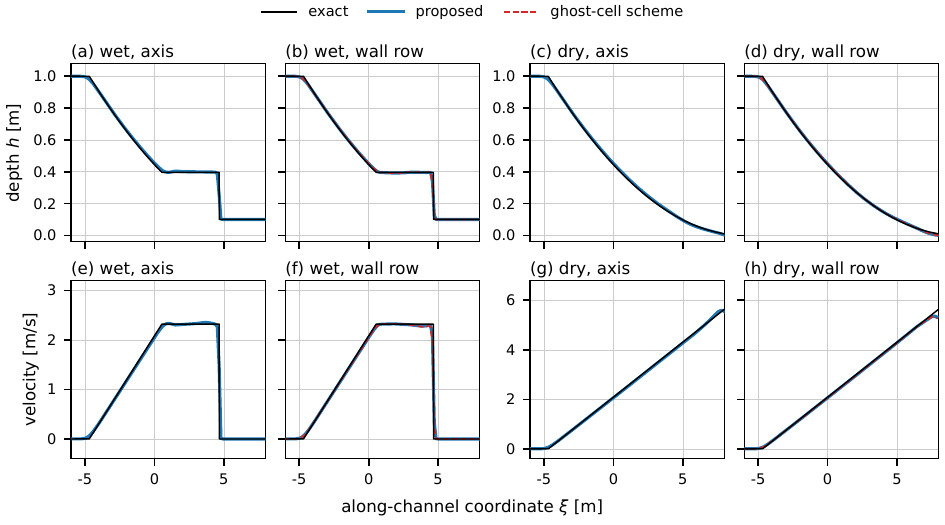}
\caption{Dam break in a channel rotated by $45^\circ$, $t=1.5$~s: depth (top) and along-channel
velocity (bottom) on the wet bed (a, b, e, f) and on the dry bed (c, d, g, h), on the channel axis and in the row
of fluid cells next to the wall, with the exact solution (black). Proposed scheme (thick blue) and ghost-cell
scheme (red, dashed). For the velocity, cells shallower than 1~mm are omitted.}\label{fig:channel}
\end{figure}

Figure~\ref{fig:channel} compares the computed depth and along-channel velocity with the
exact solution. The mean errors of the depth, the velocity and the discharge are the same for the two schemes to
within 4\% (Table~\ref{tab:pchannel}). The cut-cell continuity equation thus removes the volume error without
changing the accuracy. The cross-channel velocity next to the wall stays below 0.03~m/s, and no deceleration of
the flow along the wall is visible. With the ghost-cell scheme, the errors of the eight constructions agree to within 5\% in the
depth and the velocity and to within 11\% in the discharge, whereas their volume error varies by a factor of about
three.

\begin{figure}[!htb]
\centering
\includegraphics[width=0.85\linewidth]{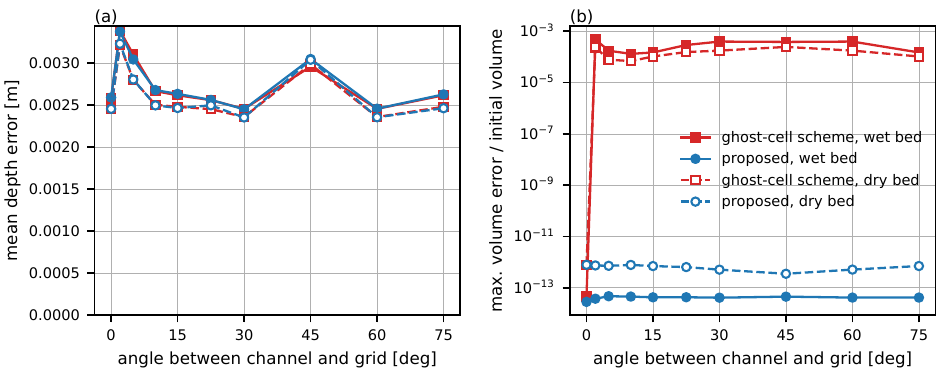}
\caption{Dam break in the channel for different angles between the channel and the grid
($\Delta x=0.093$~m for all angles, $t=1.5$~s). (a) Mean depth error against the exact solution; (b) largest
volume balance error during the run relative to the initial volume. Blue: proposed scheme; red: ghost-cell scheme. Solid lines and filled symbols: wet bed;
dashed lines and open symbols: dry bed. At $0^\circ$ the walls are parallel to the grid lines and no water crosses
them in either scheme.}\label{fig:angle}
\end{figure}

The dependence on the orientation of the grid is shown in Fig.~\ref{fig:angle}, for which the channel was
rotated by ten angles between 0 and $75^\circ$ at a fixed grid spacing and computed with the ghost-cell scheme
and the proposed scheme; the results at 60 and $75^\circ$ equal
those at 30 and $15^\circ$ by symmetry. The error of the proposed scheme lies
between $2.4\times10^{-3}$ and $3.4\times10^{-3}$~m and is largest at $2^\circ$, where the wall crosses a grid
line only every 29 cells, and at $45^\circ$; at $0^\circ$ it is $2.5\times10^{-3}$~m. The two schemes agree to
within about 3\%. The volume error of the ghost-cell scheme is between $7\times10^{-5}$ and
$5\times10^{-4}$ of the volume at every angle other than $0^\circ$, and that of the proposed scheme is below
$10^{-12}$.

\FloatBarrier
\subsection{Oblique jump}\label{sec:oblique}

\begin{figure}[!htb]
\centering
\includegraphics[width=0.45\linewidth]{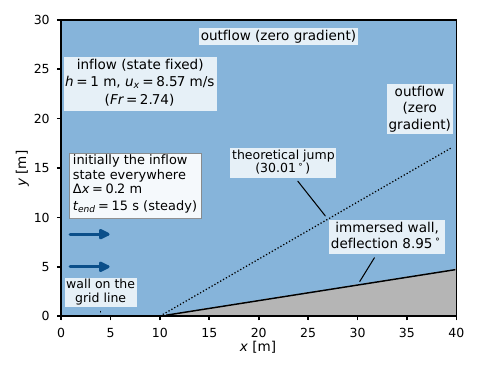}\hfill
\includegraphics[width=0.50\linewidth]{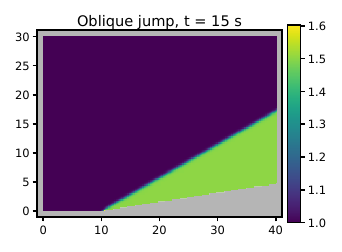}
\caption{Oblique jump. Left: set-up (gray: solid; blue: water at $t=0$). Right: computed water level $\eta$ [m] in
the steady state with the proposed scheme.}\label{fig:poblique}
\end{figure}

A supercritical stream ($h=1$~m, $u_x=8.57$~m/s, Froude number $Fr=u_x/\sqrt{gh}=2.74$) is deflected by a wall inclined by $8.95^\circ$, which
produces an oblique hydraulic jump (Fig.~\ref{fig:poblique}); this is a common benchmark (e.g.\ \cite{seo2014}).
The bed is horizontal. The wall is straight for the first 10~m and is then deflected inward. At the inflow edge
the depth and the velocity are prescribed. Theory gives a
jump angle of $30.01^\circ$ and
a depth of 1.4999~m behind the jump, i.e.\ in the deeper region next to the wall in Fig.~\ref{fig:poblique}.
Two errors are evaluated in the steady state, at $t=15$~s, for
$22<x<36$~m. The first is the relative error of the depth behind the jump, taken as the mean
depth within $3\,\Delta x$ of the wall. The second is the error of the jump angle: in each column of cells the
jump is located where the depth equals the mean of the depths ahead of and behind it, 1.25~m, and a straight line
is fitted to these positions.

\begin{table}[!htb]
\centering\small
\caption{Oblique jump: volume balance error, relative error of the depth behind the jump and error of the jump angle
($\Delta x=0.2$~m).}\label{tab:poblique}
\setlength{\tabcolsep}{2.5pt}
\begin{tabular}{lccc}
\toprule
Scheme & balance [\%] & depth error (rel.) & angle error [$^\circ$] \\
\midrule
Ghost cell & 1.55 & $2.2\times10^{-5}$ & $4.0\times10^{-3}$ \\
Cut-cell continuity (proposed) & $2.5\times10^{-11}$ & $7.7\times10^{-7}$ & $4.6\times10^{-3}$ \\
\midrule
Ghost cell, eight constructions (0 diverged) & 1.55--3.3 & $1.7\times10^{-5}$--$2.3\times10^{-4}$ & $2.7\times10^{-3}$--$5.4\times10^{-3}$ \\
\bottomrule
\end{tabular}
\end{table}

\begin{figure}[!htb]
\centering
\includegraphics[width=0.72\linewidth]{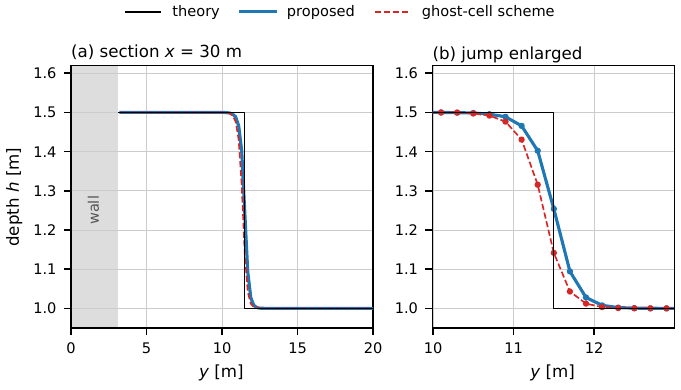}
\caption{Oblique jump, steady state, $\Delta x=0.2$~m.
(a) Depth along the grid line $x\approx30$~m; black: theory; the gray band is the solid. (b) Enlargement of (a)
at the jump; symbols are cell values. Proposed scheme (thick blue) and ghost-cell scheme (red, dashed).}\label{fig:oblique}
\end{figure}

All schemes reproduce the jump, with an error of the angle below $0.01^\circ$
(Table~\ref{tab:poblique}, Fig.~\ref{fig:oblique}). With the ghost-cell scheme, the balance error of 1.5\% is the leak through the
inclined wall behind the jump; ahead of the jump, where the flow is uniform along the wall, the reflected ghost state equals the fluid state
and nothing leaks.

Along the wall, the depth of the proposed scheme is as uniform as that of the ghost-cell scheme: in both, it deviates from the theoretical value by less than 0.001~m behind the corner region.
This requires accurate apertures; with apertures obtained by counting sample
points, the depth along the wall showed a sawtooth pattern. On the grid line of Fig.~\ref{fig:oblique}, the jump of
the proposed scheme lies $+0.010$~m ($+0.05\,\Delta x$) from its theoretical position, compared with $-0.12$~m
($-0.6\,\Delta x$) for the ghost-cell scheme (Fig.~\ref{fig:oblique}b), whose volume balance does not close (1.5\% of the volume
in the domain).

\FloatBarrier
\subsection{Vortex}\label{sec:vortex}

\begin{figure}[!htb]
\centering
\includegraphics[width=0.40\linewidth]{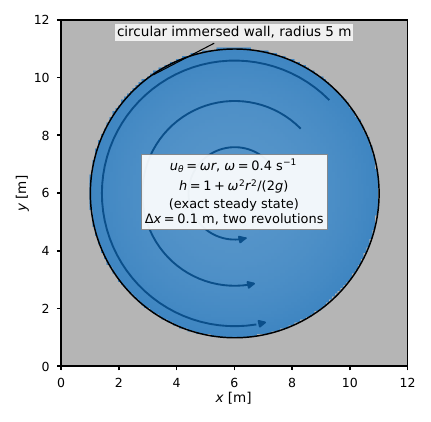}\hspace{0.04\linewidth}
\raisebox{0.012\linewidth}{\includegraphics[width=0.46\linewidth]{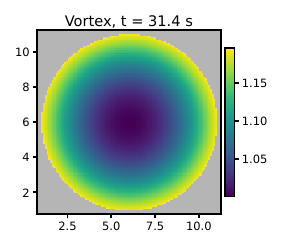}}
\caption{Vortex. Left: set-up (gray: solid; blue: water at $t=0$, darker where deeper). Right: water level $\eta$ [m] after two revolutions, computed with the proposed scheme.}\label{fig:pvortex}
\end{figure}

Rigid rotation $u_\theta=\omega r$ with depth $h=h_0+\omega^2r^2/(2g)$ is an exact steady solution in a
circular basin with a slip wall (Fig.~\ref{fig:pvortex}); here the radius is 5~m, $h_0=1$~m and
$\omega=0.4$~s$^{-1}$, and the bed is horizontal. The exact solution is imposed as the initial condition, and
the flow is computed for two revolutions without friction or forcing; any change of the flow is an error of the
scheme. The flow along the curved wall is fast (2~m/s) and
the water level has a normal gradient at the wall, so the problem is a severe test of the wall treatment. The errors
are the mean error of the speed and the loss of angular momentum after two revolutions; a circular wall exerts
no torque about its center, so angular momentum should be conserved.

\begin{table}[!htb]
\centering\small
\caption{Vortex: volume balance error, mean speed error and loss of angular momentum after two revolutions ($\Delta x=0.1$~m).}\label{tab:pvortex}
\setlength{\tabcolsep}{4pt}
\begin{tabular}{lccc}
\toprule
Scheme & balance [\%] & speed error [m/s] & ang.\ mom.\ loss [\%] \\
\midrule
Ghost cell & 4.68 & 0.105 & 13.3 \\
Cut-cell continuity (proposed) & $4.4\times10^{-13}$ & 0.0666 & 6.39 \\
\midrule
Ghost cell, eight constructions (1 diverged) & 0.612--45.2 & 0.0801--0.401 & 8.46--52.2 \\
\bottomrule
\end{tabular}
\end{table}

\begin{figure}[!htb]
\centering
\includegraphics[width=\linewidth]{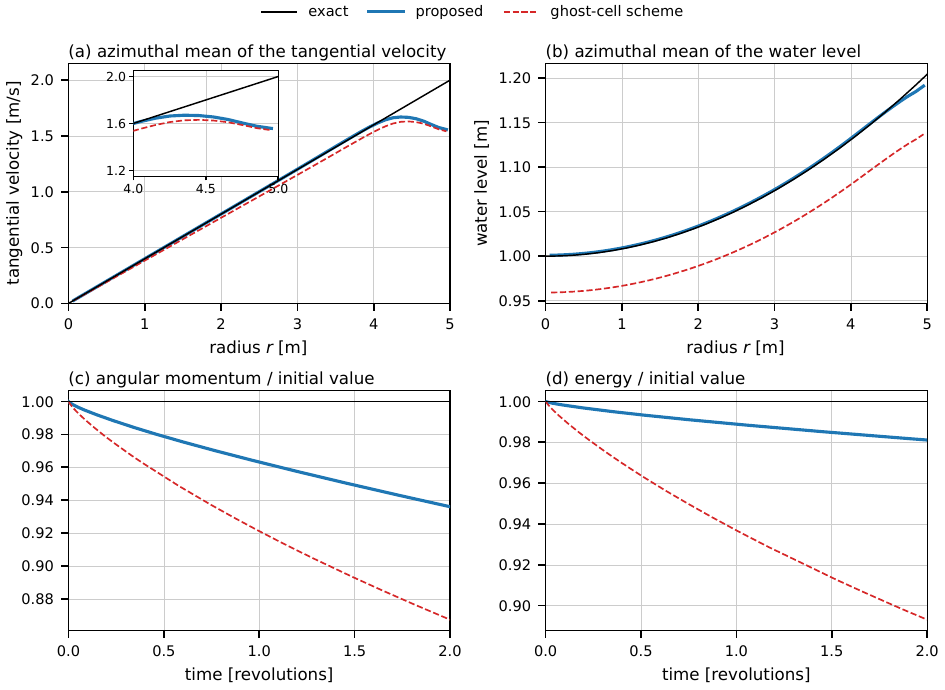}
\caption{Steady vortex after two revolutions, $\Delta x=0.1$~m. (a) Tangential velocity averaged over rings of
width $\Delta x$ against the radius, with the exact $\omega r$ (black); the inset enlarges the outer meter.
(b) Water level averaged over the same rings, with the exact solution (black). (c) Angular momentum
about the center of the basin and (d) energy, divided by their initial values, against time. Proposed scheme (thick blue), ghost-cell scheme (red, dashed). The energy loss of the
ghost-cell scheme includes the energy of the water that is lost.}\label{fig:vortex}
\end{figure}

The ghost-cell scheme loses 4.7\% of the volume and 13\% of the angular momentum in two
revolutions (Table~\ref{tab:pvortex}, Fig.~\ref{fig:vortex}); the seven constructions that completed the run lose between 0.6 and 45\% of
the volume. The volume loss with the baseline interpolation is 8.8, 4.7 and 2.5\% for $\Delta x=0.2$, 0.1 and
0.05~m, i.e.\ it
decreases with an observed order of 0.9. The cut-cell continuity equation removes the volume error and halves the loss of angular momentum
(Table~\ref{tab:pvortex}). It does not repair
the loss of tangential momentum at the wall (Fig.~\ref{fig:vortex}a): in the cells within $\Delta x$ of the
wall, the tangential velocity is 21\% below the exact value with the cut-cell continuity equation and 22\%
with the ghost-cell scheme, and the angular momentum is still decreasing by about 3\%
per revolution at the end of the computation. In the interior ($r<4$~m) the largest velocity error is
0.007~m/s, compared with 0.065~m/s with the ghost-cell scheme. The water level averaged over rings stays within
0.011~m of the exact solution, whereas it falls by up to 0.066~m with the ghost-cell scheme, which loses water
(Fig.~\ref{fig:vortex}b). The energy decreases by 1.9\% in two revolutions, compared with 11\% for the ghost-cell
scheme (Fig.~\ref{fig:vortex}d). The ghost-cell treatment of the momentum is first-order accurate, neglects the curvature of the wall and does not
conserve momentum. The vortex is the problem in which these weaknesses matter most.

\FloatBarrier
\subsection{Network}\label{sec:network}

\begin{figure}[!htb]
\centering
\raisebox{0.02\linewidth}{\includegraphics[width=0.39\linewidth]{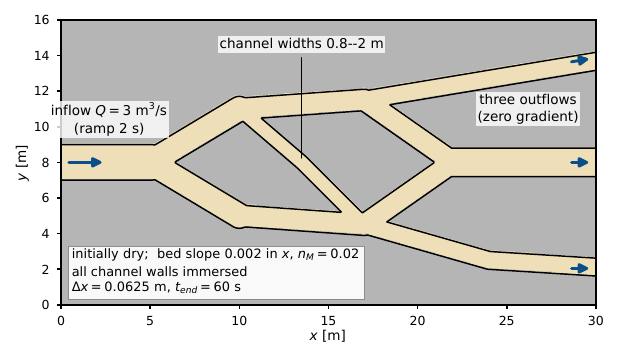}}\hfill
\includegraphics[width=0.59\linewidth]{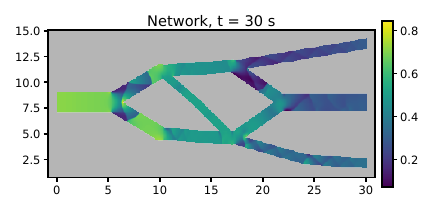}
\caption{Network. Left: set-up (gray: solid; blue: water at $t=0$, darker where deeper; sand: dry bed). Right: water level $\eta$ [m] at $t=30$~s, computed with the proposed scheme.}\label{fig:pnetwork}
\end{figure}

A discharge of 3~m$^3$/s, reached linearly in 2~s, enters a dry network of channels, 0.8--2~m wide, on a slope
of 0.002 with $n_M=0.02$~s\,m$^{-1/3}$; the network divides,
reconnects and leaves the domain through three outlets. As the problem serves to show robustness, the layout is
given by Fig.~\ref{fig:pnetwork} only. All walls are immersed and
most are oblique. The wetting front runs along them, and once the network has filled, the flow along the walls
is fast. There is no reference solution; the problem tests robustness and the volume balance.

\begin{table}[!htb]
\centering\small
\caption{Network: volume balance error at $t=60$~s ($\Delta x=0.0625$~m).}\label{tab:pnetwork}
\begin{tabular}{lc}
\toprule
Scheme & balance [\%] \\
\midrule
Ghost cell & 33.5 \\
Cut-cell continuity (proposed) & $2.3\times10^{-10}$ \\
\midrule
Ghost cell, eight constructions (5 diverged) & 24.4--34.6 \\
\bottomrule
\end{tabular}
\end{table}

This problem is the most demanding in terms of robustness (Table~\ref{tab:pnetwork}). With the ghost-cell scheme,
five of the eight constructions diverge: the constrained bilinear interpolation, MLS1, MLS2, WLS1 and WLS2.
Those that complete the run have balance
errors of 24--35\% of the volume (baseline: 34\%). This is the ordinary wall
leak. Since a uniform flow along a straight wall does not leak, the leak must arise where
the flow is not uniform along the walls: at the junctions and bends and at the advancing front. The proposed scheme conserves the volume.

\FloatBarrier
\subsection{Dam break in a channel with a $45^\circ$ bend}\label{sec:bend45}

\begin{figure}[!htb]
\centering
\includegraphics[width=0.85\linewidth]{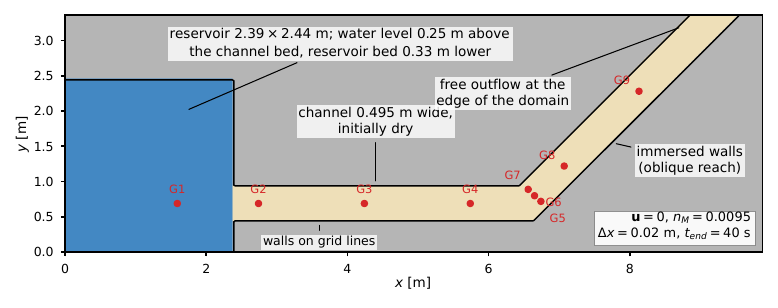}
\caption{$45^\circ$ bend: set-up (gray: solid; blue: water at $t=0$, darker where deeper; sand: dry bed; red
dots: gauges).}\label{fig:pbend45}
\end{figure}

The laboratory case is the CADAM dam-break flow in a channel with a $45^\circ$ bend
(Fig.~\ref{fig:pbend45}). The
dimensions, initial conditions and gauge positions are taken from Lai and Khan~\cite{lai2011}: a
reservoir of $2.39\times2.44$~m, a channel 0.495~m wide with reaches of 4.25~m before and 4.15~m after the bend
(outer wall), a water level in the reservoir 0.25~m above the channel bed, an initially dry channel and a Manning coefficient
of $n_M=0.0095$~s\,m$^{-1/3}$; the walls are slip walls. Lai and Khan~\cite{lai2011} describe the bed as horizontal, whereas Zhou et al.~\cite{zhou2004} place the
reservoir bed 0.33~m below the channel bed. With this step, the computed depths
agreed better with the measurements than with a horizontal bed, and the results shown use the stepped bed. The reservoir and the first reach are parallel to the grid lines, but only two of their walls coincide with
cell faces (one reservoir wall passes through cell centers); the second reach is oblique to the grid and is continued beyond the exit section of the experiment to the edge
of the domain, where the outflow is a free overfall. The measurements
at the nine gauges and the discontinuous Galerkin solution of Lai and Khan~\cite{lai2011} were digitized from Figs.~18--26 of
that paper; the grid spacing is $\Delta x=0.02$~m.

\begin{figure}[!htb]
\centering
\includegraphics[width=0.93\linewidth]{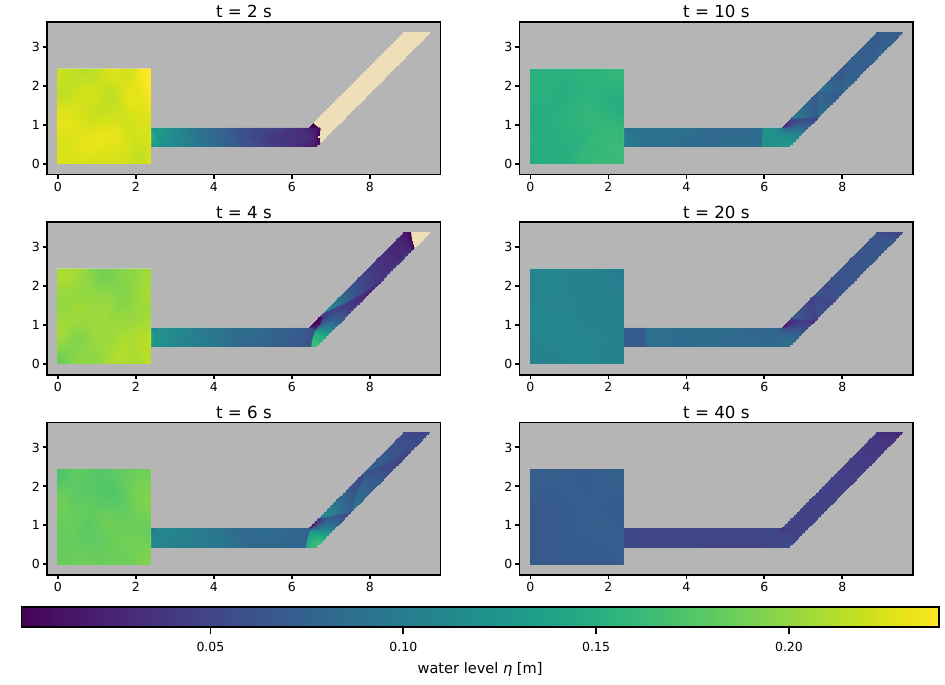}
\caption{$45^\circ$ bend: water level $\eta$ [m] above the channel bed at six times, computed with the proposed scheme
(sand: dry bed). The front reaches the bend at about 2~s and is reflected there; the reflected bore travels
upstream and is at $x\approx6$~m at 10~s and at $x\approx3$~m at 20~s.}\label{fig:bendseries}
\end{figure}

\begin{figure}[!htb]
\centering
\includegraphics[width=\linewidth]{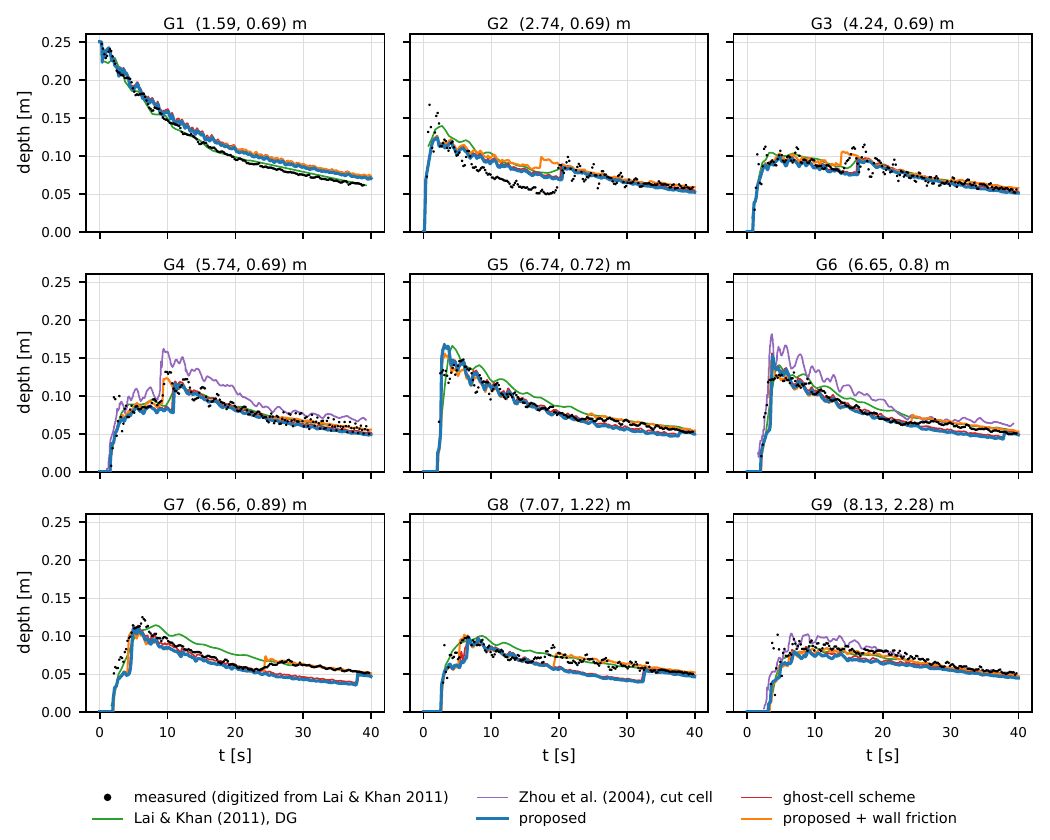}
\caption{Dam break in a channel with a $45^\circ$ bend: depth above the channel bed at the nine gauges
(coordinates in m). The measurements and the solution of Lai and Khan~\cite{lai2011} were digitized from that paper, and
the solution of Zhou et al.~\cite{zhou2004} (G4, G6 and G9 only) from theirs. Orange: proposed scheme with the
wall friction term described in the text. The
curve of the ghost-cell scheme is hidden by that of the proposed scheme where the two coincide.}\label{fig:bend45}
\end{figure}

Figure~\ref{fig:bendseries} shows the computed water level at six times. The arrival of the front, the bore
reflected at the bend and the subsequent decay are reproduced at all nine
gauges (Fig.~\ref{fig:bend45}). The mean absolute difference from the measured depth (computed record interpolated to the times of the
digitized measurements), averaged over the gauges,
is 0.0085~m for the ghost-cell scheme and 0.0094~m with the cut-cell continuity equation (maxima over the
gauges: 0.011 and 0.013~m; Table~\ref{tab:pbend45}), compared with 0.0078~m for the solution of Lai and Khan~\cite{lai2011}. With a horizontal
reservoir bed, the difference for the proposed scheme is 0.0100~m. Zhou et al.~\cite{zhou2004} computed the
experiment with a Cartesian cut-cell method ($\Delta x=0.05$~m) and show three gauges: at G4, G6 and G9 the mean
difference of their solution is 0.018, 0.016 and 0.007~m, against 0.009, 0.008 and 0.011~m for the proposed
scheme. At the two gauges behind the bend (G7 and G8), the computed depth is 0.02--0.03~m below the measured one from 24~s until it rises abruptly by about 0.013~m (at 32~s at G8
and at 38~s at G7), when a jump that stands downstream of the bend moves upstream across the
gauges. The measurements show a smaller rise much earlier (at about 18~s at G8 and 25~s at G7); the upstream motion
of the jump is thus delayed in the computation. Both schemes produce the delay.

The delay is largely explained by the friction of the side walls, which the computations above do not include. To show
this, the proposed scheme was run with a wall friction term added to the bed friction. The walls are assumed to
exert the same Manning-type shear stress as the bed, with a coefficient $n_w$ in place of $n_M$. The force on the two walls, each wetted over the depth $h$, is distributed over the local width $b$
(0.495~m in the channel and 2.44~m in the reservoir), which replaces $n_M^2$ in the friction term (Section~\ref{sec:base}) by
$n_M^2+n_w^2\,2h/b$. The wall coefficient is $n_w=0.0195$~s\,m$^{-1/3}$, the value given by Lai and Khan~\cite{lai2011}. The result is shown in
orange in Fig.~\ref{fig:bend45}. The
rises then occur at 19~s at G8 and 24~s at G7, the bore reflected at the bend reaches G4 at 9.4~s as measured
(10.9~s without wall friction), and the mean difference from the measurements decreases by 25--55\% at G4 to G9
and from 0.0094 to 0.0072~m over the nine gauges. Brufau and Garc\'{\i}a-Navarro~\cite{brufau2000}, who computed
the same experiment, report that wall friction advances the reflected front as well: without it, their front
arrived 3~s late at G3.

In the first reach the wall friction makes the agreement worse: the reflected bore arrives about 3~s early at G3
and G2 (14.0 and 17.2~s, against 16.9 and 19.8~s measured and 16.3 and 20.5~s without wall friction). Runs with the
term restricted to one reach showed that the two reaches act independently; with wall friction in the bend and
the second reach only, the first reach is unchanged and the mean difference is 0.0068~m. The early arrival is not
a numerical effect: halving the time step or changing the grid spacing to 0.04 or 0.01~m changed these times by
at most 1.2~s, and a one-dimensional friction law with a composite roughness and the hydraulic radius changed
them by at most 0.5~s. A single wall coefficient thus does not fit both reaches. Between 5 and 20~s the computed
depth at G2, 0.35~m downstream of the gate, exceeds the measured one in all computations, including that of Lai
and Khan, which may indicate that the flow leaving the reservoir is not computed well there. The flow contracts
from the reservoir into the channel at this place and may be three-dimensional and strongly turbulent, whereas
the computations are depth-averaged and include no turbulence model.

The two schemes give almost the same gauge records, so that this experiment does not discriminate between the
wall treatments in the depth. They differ in the volume balance: 3.2\% for the ghost-cell scheme and
$1.5\times10^{-10}$\% for the proposed scheme.

\begin{table}[!htb]
\centering\small
\caption{$45^\circ$ bend: volume balance error and time-averaged absolute difference from the measured depth, given as the mean and the maximum over the nine gauges ($\Delta x=0.02$~m).}\label{tab:pbend45}
\setlength{\tabcolsep}{3pt}
\begin{tabular}{lccc}
\toprule
Scheme & balance [\%] & mean diff.\ [m] & max.\ over gauges [m] \\
\midrule
Ghost cell & 3.18 & $8.5\times10^{-3}$ & 0.0112 \\
Cut-cell continuity (proposed) & $1.5\times10^{-10}$ & $9.4\times10^{-3}$ & 0.0126 \\
\midrule
Ghost cell, eight constructions (2 diverged) & 1.17--3.18 & $8.5\times10^{-3}$--$9.5\times10^{-3}$ & 0.0111--0.0117 \\
\bottomrule
\end{tabular}
\end{table}

\FloatBarrier
\FloatBarrier
\section{Summary of the comparison and choice of the scheme}\label{sec:summary}

The results of the individual problems are now brought together to decide which treatment of the immersed
wall is preferable.

\FloatBarrier
\subsection{Robustness}\label{sec:first}

Table~\ref{tab:eval1} lists the problems in which the computation diverged, for the eight constructions of the
ghost values and for two schemes: the ghost-cell scheme and the proposed scheme. The three constructions whose
weights are non-negative (the baseline, the fixed probe distance and IDW) completed all seven problems with both
schemes. With the ghost-cell scheme, the five whose weights can be negative all failed in the network, MLS1 also
in the bend, and MLS2 in two further problems; negative weights, applied separately to depth and discharge, can produce large
velocities in shallow water. Within the proposed scheme, the constrained bilinear interpolation and WLS1
completed all seven problems as well, whereas MLS1, MLS2 and WLS2 failed in one to three. Raising the degree from
one to two did not help: with either scheme MLS2 failed in three problems, against two (ghost-cell scheme) and
one (proposed scheme) for MLS1, and within the
proposed scheme WLS2 failed in two problems whereas WLS1 completed all seven. MLS1, MLS2 and WLS2 thus remain
unreliable within the proposed scheme.

\begin{table}[!htb]
\centering\small
\caption{Problems in which the computation diverged, out of the seven problems, for the eight constructions of
the ghost values with the ghost-cell scheme and within the proposed scheme. A dash means
that all seven problems were completed. Weights: $+$ non-negative, $\pm$ can be negative
(Table~\ref{tab:methods}).}\label{tab:eval1}
\begin{tabular}{lccc}
\toprule
Construction & Weights & Ghost-cell scheme & Proposed scheme \\
\midrule
G-bilinear (baseline) & $+$ & --- & --- \\
G-bilinear, fixed probe & $+$ & --- & --- \\
G-IDW & $+$ & --- & --- \\
G-constrained bilinear & $\pm$ & network & --- \\
G-WLS1 & $\pm$ & network & --- \\
G-MLS1 & $\pm$ & network, bend & network \\
G-WLS2 & $\pm$ & network & network, bend \\
G-MLS2 & $\pm$ & vortex, network, bend & vortex, network, bend \\
\bottomrule
\end{tabular}
\end{table}

\FloatBarrier
\subsection{Volume conservation}\label{sec:mass}

\begin{figure}[!htb]
\centering
\includegraphics[width=\linewidth]{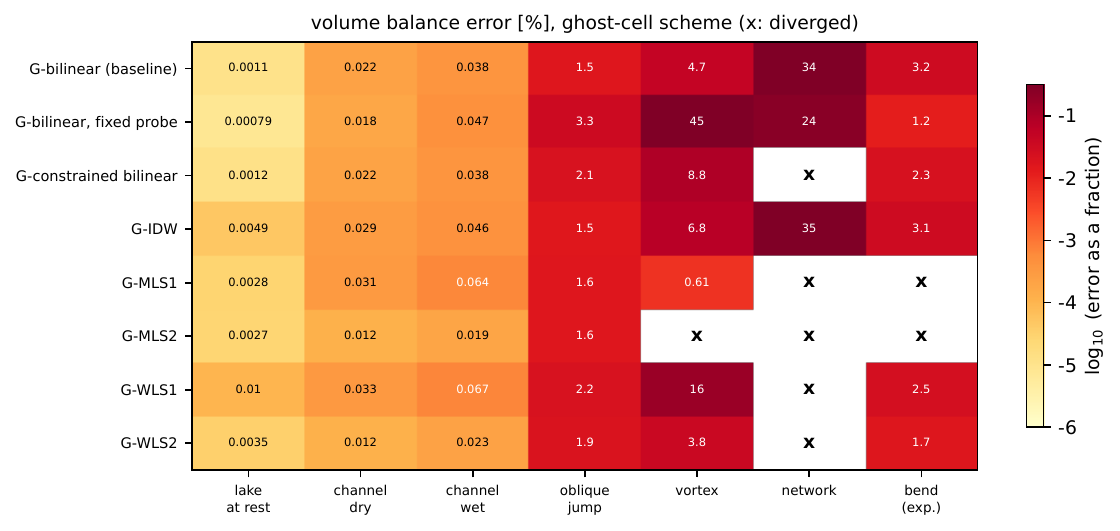}
\caption{Volume balance error [\%] of the eight constructions of the ghost values in the seven problems
(ghost-cell scheme; x marks diverged runs).}\label{fig:eval1}
\end{figure}

Every ghost-cell variant leaks mass through the wall, and no interpolation removes the leak
(Fig.~\ref{fig:eval1}). In the
vortex the leak decreases with an observed order of 0.9
(Section~\ref{sec:vortex}). With the baseline interpolation it is
small in the channel (0.02--0.04\%) and large where the flow is
fast and not uniform along the walls: 1.5\% in the oblique jump, 4.7\% in the vortex, 3.2\%
in the $45^\circ$ bend, and 34\% in the network
(first row of Fig.~\ref{fig:eval1}). In a steady flow with open boundaries the leak is a loss of discharge through the walls; the balance error
quoted for the oblique jump is this loss accumulated over the run, relative to the volume in the domain. A uniform flow along a straight wall does not leak through the wall.

The cut-cell continuity equation conserves the volume to within $3\times10^{-12}$ of its value at the
end of every problem (Table~\ref{tab:eval2}).

In the proposed scheme no negative depth occurred in the seven problems. The clipping of depths below
$\varepsilon$ accounts for the balance error on dry beds (in the channel, for example, the clipped volume
is $6.6\times10^{-12}$~m$^3$ and the balance error $6.9\times10^{-12}$~m$^3$); on wet beds nothing is clipped, and the balance
closes to round-off. The volume moved by the redistribution, accumulated over a run, is between 0.04 times
(channel) and 12 times (network) the water held in the cut cells: the redistribution is an essential part of the
mass balance near the wall and not a small correction.

\FloatBarrier
\subsection{Accuracy}\label{sec:second}

\begin{figure}[!htb]
\centering
\includegraphics[width=\linewidth]{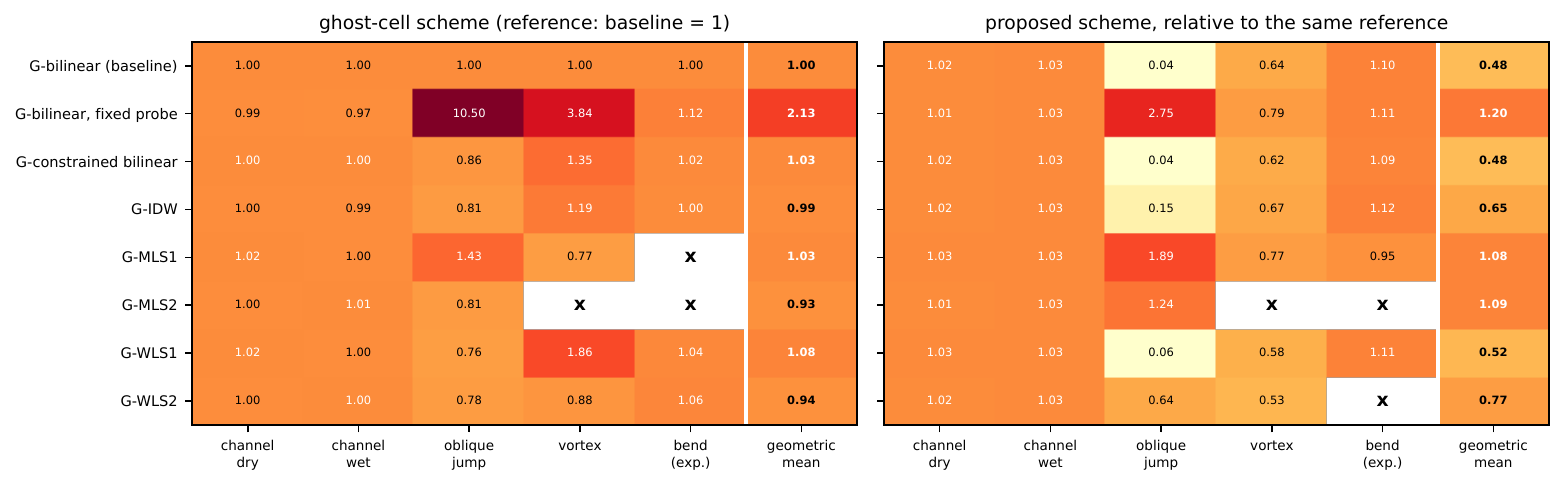}
\caption{Error of the eight constructions of the ghost values in the problems with a reference solution or
measurements, divided by the error of the ghost-cell scheme with the baseline interpolation (top row of the left
panel, equal to 1). Left: ghost-cell scheme. Right: proposed scheme, i.e.\ cut-cell continuity equation with the
dry-cell treatment, divided by the same reference; its top row is the proposed scheme with the baseline
interpolation. The last column of each panel is the geometric mean over the problems of the row. x marks diverged
runs, which are excluded from the mean.}\label{fig:eval2}
\end{figure}

In the problems with a reference, the error of the other constructions with the ghost-cell scheme is within 10\%
of that of the baseline (error ratios 0.93--1.08 in the geometric mean over these problems, last column of
Fig.~\ref{fig:eval2}, left), except for the fixed probe distance,
which doubles it;
in these problems, most of which are dominated by the
error of the interior scheme at discontinuities, a higher interpolation degree brings no benefit.
The fixed probe distance was tested as a way of keeping the ghost cell itself out of the stencil: at
$1.5\Delta x$ from a straight wall, all four nodes around the interpolation point lie in the fluid. It had the
opposite effect: the error increased by a factor of 10 in the oblique jump and of 4 in the vortex, and the leak
in the vortex rose from 4.7 to 45\%.

Table~\ref{tab:eval2} and Fig.~\ref{fig:eval2} compare the two schemes.
Relative to the ghost-cell scheme, the error of the
proposed scheme is 36\% smaller in the vortex, 2--3\% larger in the
channel (ratios 1.02 and 1.03) and 10\% larger in the $45^\circ$ bend, where the difference from the
measurements is dominated by effects other than the wall treatment (Section~\ref{sec:bend45}). In the oblique
jump both errors are very small ($2\times10^{-5}$ and $8\times10^{-7}$ relative
error of the depth; Table~\ref{tab:poblique}).

Within the proposed scheme, the errors of the constrained bilinear interpolation and of WLS1 are within 9\% of
that of the baseline in the geometric mean over the five problems of Table~\ref{tab:eval2}; IDW and the fixed probe
distance are less accurate, mainly in the oblique jump (Fig.~\ref{fig:eval2}, right). With the mass conserved by
the continuity equation, the choice among the robust interpolations thus has little influence on the accuracy.

\begin{table}[!htb]
\centering\small
\caption{Baseline interpolation. First row: largest volume balance error in the seven problems. Other rows: errors in the
problems with a reference (channel: mean depth error [m]; vortex: mean speed error [m/s];
oblique jump: relative error of the depth behind the jump;
$45^\circ$ bend: mean absolute difference from the measured depth at the gauges [m]).}
\label{tab:eval2}
\setlength{\tabcolsep}{4pt}
\begin{tabular}{lcc}
\toprule
 & ghost-cell scheme & proposed scheme \\
\midrule
balance, maximum [\%] & 33.5 & $2.3\times10^{-10}$ \\
\midrule
channel, dry & $3.0\times10^{-3}$ & $3.0\times10^{-3}$ \\
channel, wet & $3.0\times10^{-3}$ & $3.1\times10^{-3}$ \\
oblique jump & $2.2\times10^{-5}$ & $7.7\times10^{-7}$ \\
vortex & 0.105 & 0.0666 \\
$45^\circ$ bend & $8.5\times10^{-3}$ & $9.4\times10^{-3}$ \\
\bottomrule
\end{tabular}
\end{table}

\begin{figure}[!htb]
\centering
\includegraphics[width=0.75\linewidth]{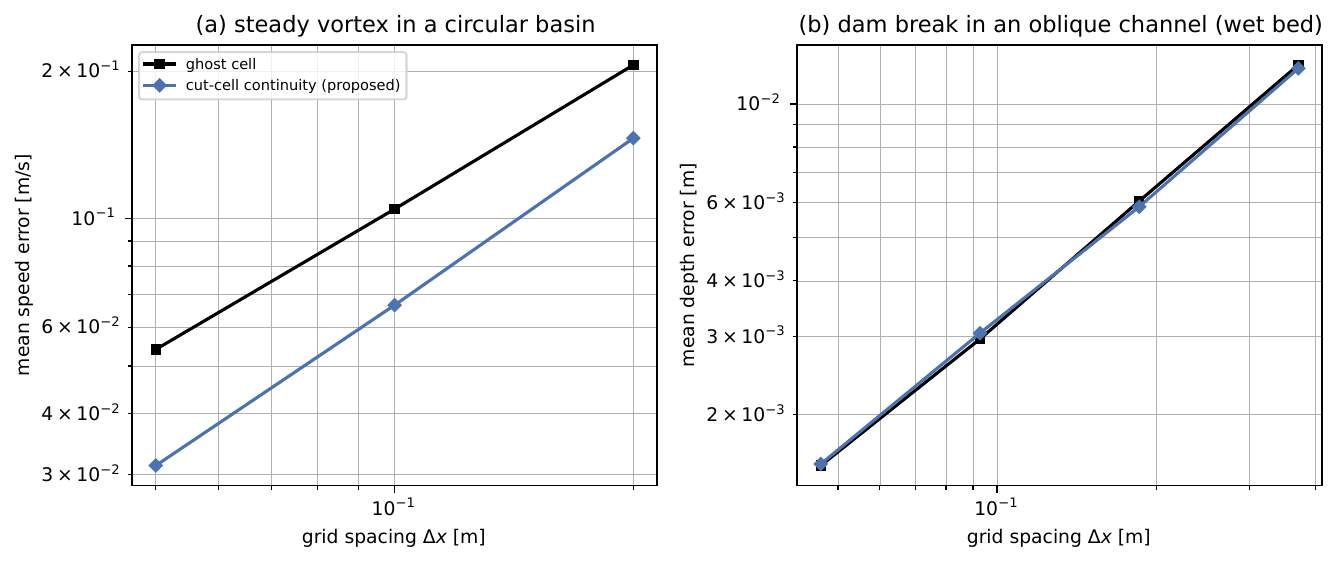}
\caption{Convergence. (a) Steady vortex. (b) Dam break in a channel rotated
by $45^\circ$.}\label{fig:order}
\end{figure}

Table~\ref{tab:unsplit} and Fig.~\ref{fig:order} show the convergence. In the channel, the errors of the two schemes decrease at the
same rate, which is that of the interior scheme at the discontinuities. In the vortex, the error of
both schemes decreases only at first order: the wall conditions imposed through the ghost cells are
first order (Section~\ref{sec:bilinear}), and on a curved wall the condition of zero normal gradient of the water level
neglects the centrifugal force. No curvature correction is included. This limits the accuracy there. The cut-cell
continuity equation reduces the error by 29 to 43\% but does not change the order.

\begin{table}[!htb]
\centering\footnotesize
\caption{Convergence.
Vortex: mean speed error [m/s] and loss of angular momentum [\%] for $\Delta x=0.2$, 0.1, 0.05~m.
Channel: mean depth error [$10^{-3}$~m], wet bed, $\Delta x=0.37$, 0.19, 0.093, 0.046~m.}\label{tab:unsplit}
\setlength{\tabcolsep}{4pt}
\begin{tabular}{lccc}
\toprule
Scheme & vortex, speed error & vortex, ang.\ mom. & channel \\
\midrule

Ghost cell & 0.206 / 0.105 / 0.054 & 25 / 13 / 7.1 & 12.2 / 6.04 / 2.96 / 1.54 \\

Cut-cell continuity (proposed) & 0.146 / 0.067 / 0.031 & 14 / 6.4 / 3.1 & 12.0 / 5.87 / 3.05 / 1.55 \\
\bottomrule
\end{tabular}
\end{table}

\FloatBarrier
\subsection{Wetting and drying}\label{sec:dry}

On a sloping bed with a shoreline at the wall, the exclusion of dry nodes from the interpolation of the water level
(Section~\ref{sec:drydef}) is necessary: without it, every ghost-cell variant sets the lake at rest in motion
(Section~\ref{sec:lake}). With both parts of the dry-cell treatment (Section~\ref{sec:drydef}), the lake stayed at rest to
round-off in the tests, with or without the cut-cell continuity equation. Added to the ghost-cell scheme without the cut-cell continuity equation, the treatment changes nothing on wet
beds, where it is not active, and little on dry beds: in the channel on a dry bed the mean depth error changes
from $2.99\times10^{-3}$ to $3.06\times10^{-3}$~m, and in the $45^\circ$ bend the balance error and the
differences from the measurements are the same to three digits. The treatment is no substitute
for the conservation of mass: the balance error in the network is a wall leak and is 33.5\% with and without it.

\FloatBarrier
\subsection{Choice of the scheme}

\paragraph{Which construction of the ghost values?}
The renormalized bilinear interpolation at the image point is among the most robust of the eight constructions; its accuracy and volume error lie within the range of the others. It is also the easiest to implement. Its weights
are the four bilinear weights at the image point, of which those of the non-fluid nodes are dropped and the rest
are rescaled; no system of equations is solved, and the same rule covers every arrangement of fluid, solid and dry
nodes. The constrained bilinear interpolation and the MLS and WLS variants require a linear system to be solved
for each ghost cell (Table~\ref{tab:methods}). In the constrained bilinear interpolation this is a $4\times4$
system for the water level and an $8\times8$ system for the discharge, whose two components are coupled by the
wall conditions. In addition, these variants need a substitute construction for the ghost cells in which the
system cannot be solved. MLS1, MLS2 and WLS2 have a smaller leak
with the ghost-cell scheme, but each of them failed in one or more problems with either scheme
(Table~\ref{tab:eval1}). Moving the interpolation point away from the wall and degree-two interpolation
are not recommended on the basis of these tests.

\paragraph{Why the cut-cell continuity equation?}
The
cut-cell continuity equation removes the leak with an accuracy comparable to that of the ghost-cell scheme
and needs only the area fractions and
apertures in addition to what the ghost-cell scheme already uses, but it requires an unsplit update.

\paragraph{What is retained by using cut cells for the continuity equation only?}
A full cut-cell method conserves mass and momentum. For the SWE with a variable bed it requires (i) fluxes and
gradients on the cut cells, (ii) a pressure force on the solid face of each cut cell, (iii) merging, or another
stabilization, of small cells for all conserved variables, and (iv) a hydrostatic reconstruction on partially dry cut
cells and on their solid faces \cite{causon2000,an2012}. The proposed scheme takes from the cut-cell method only what
is needed to conserve mass, and most of the advantages of the ghost-cell method are retained:
\begin{itemize}
\item The momentum equations are solved as in the interior: no wall pressure and no momentum flux on cut cells
are needed.
\item No cells are merged; the bound $\Lambda_{\min}$ and the redistribution act on the depth only.
\item The second-order reconstruction of the interior is used next to the wall.
\item Wetting and drying at the wall is handled by omitting the dry nodes from the interpolation.
\end{itemize}

The proposed scheme, i.e.\ the unsplit ghost-cell scheme with bilinear interpolation at the image
point, renormalized over the nodes that are neither solid nor dry, and with the cut-cell continuity equation, is
therefore adopted: it conserved the volume and kept the accuracy of the ghost-cell
scheme (errors between 36\% smaller and 3\% larger in the problems with an exact reference, the oblique jump aside, where both errors are negligible).
It is not the more accurate scheme in every problem: in the channel and the $45^\circ$ bend the ghost-cell
scheme is slightly more accurate. The strength of the proposed scheme lies in the overall result: its
accuracy does not deteriorate markedly in any of the problems, and the volume is conserved in all of them.
Where the conservation of momentum at the wall or second-order accuracy on curved walls is essential, a cut-cell
treatment of the momentum equations with a hydrostatic reconstruction on the cut cells \cite{an2012} is the
alternative, at the cost of items (i)--(iv) above.

\FloatBarrier
\section{Demonstration: run-up through an array of cylinders}\label{sec:demo}

\begin{figure}[!htb]
\centering
\includegraphics[width=0.78\linewidth]{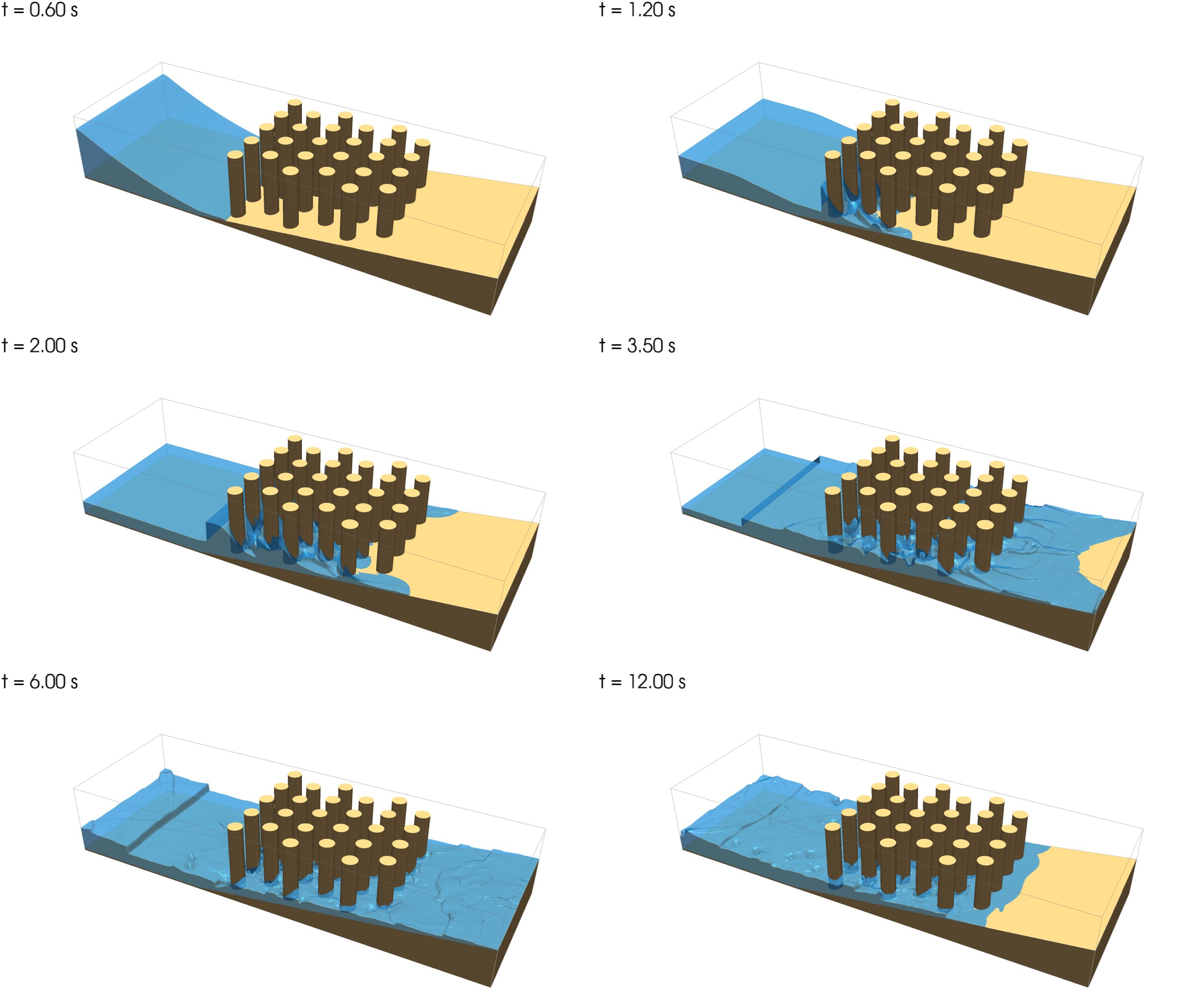}
\caption{Run-up through an array of 27 cylinders on a slope: water level and bed at six times (vertical scale
exaggerated by a factor of 1.6; dry bed in sand color).}\label{fig:cyl3d}
\end{figure}

Finally, the proposed scheme is applied to a problem that combines many curved walls with a sloping bed and
with wetting and drying, and for which no reference solution exists (Fig.~\ref{fig:cyl3d}); it illustrates the
robustness of the scheme and is not a validation. A closed tank of
$12\times4$~m has a horizontal bed for $x<4$~m and a slope of $1/12$ beyond. Twenty-seven staggered cylinders of
radius 0.2~m stand on the slope between $x=4.8$ and 9.2~m, in six rows 0.8~m apart with five and four
cylinders alternately, spaced 0.8~m across the tank. Initially, water with a depth of 1.0~m is held in
$x<2$~m, and the rest of the bed is dry. The Manning coefficient is $n_M=0.01$~s\,m$^{-1/3}$. The grid spacing is
$\Delta x=0.01$~m ($1200\times400=480\,000$ cells, of which 446\,030 are fluid cells; 20 cells per radius), and
the computation runs for 14~s.

The released water runs up the dry slope, wets the cylinders one row after the other, passes through the array
with bow waves in front of the cylinders and dry wakes behind them, reaches the end wall, and runs back down,
so that the upper part of the slope and the cylinders standing on it fall dry again. Wet--dry fronts thus cross
every cylinder twice. The tank is closed, and the volume $V$ stays constant to within $4\times10^{-12}$ of its
initial value over the 27\,318 steps of the computation. The cylinders are specified by their signed distance alone; the bed elevation
stored inside them is ignored (Section~\ref{sec:bed}).

\FloatBarrier
\section{Conclusions}\label{sec:conclusions}

A ghost-cell immersed boundary scheme for the shallow water equations on Cartesian grids has been proposed. In this scheme,
the ghost values are interpolated bilinearly at the image point; solid and dry nodes are excluded, and the
weights of the remaining nodes are renormalized. Mass is conserved by solving the continuity equation in
cut-cell form in an unsplit second-order MUSCL scheme, while the momentum equations are left unchanged. The scheme was compared, problem by problem, with
the ghost-cell scheme without these treatments and
with seven other constructions of the ghost values.
\begin{enumerate}
\item The proposed scheme conserved the volume to within $3\times10^{-12}$ of its value in all seven problems and kept
the accuracy of
the ghost-cell scheme: its error was 36\% smaller in the vortex, 2--3\% larger in the channel and 10\%
larger in the $45^\circ$ bend, and negligible for both schemes in the oblique jump.
\item It kept at rest a partially dry lake whose shoreline crosses an immersed wall, for all 89 orientations of
the wall and all water levels tested. This property, previously shown for cut cells \cite{an2012}, is observed
here and not proven.
\item Ghost-cell schemes without a mass treatment leak mass through the wall: several percent on curved walls and
tens of percent in a channel network. None of the eight interpolations removed the leak.
\item Of the eight constructions of the ghost values, the five whose weights can be negative failed in one or more
problems with the ghost-cell scheme, and three of them also within the proposed scheme. A higher degree of interpolation brought no gain in accuracy.
\item Using cut cells for the continuity equation requires no wall pressure, no momentum flux on cut cells and
no cell merging. In return, momentum is not conserved at the wall, and the wall conditions are first-order
accurate.
\end{enumerate}

\appendix
\section{MUSCL reconstruction and hydrostatic reconstruction}\label{sec:hr}

This appendix describes the base scheme of Section~\ref{sec:base} in detail. The description is given for the
$x$ direction and a row of cells $i$; the $y$ direction is treated in the same way.

\begin{figure}[t]
\centering
\includegraphics[width=\linewidth]{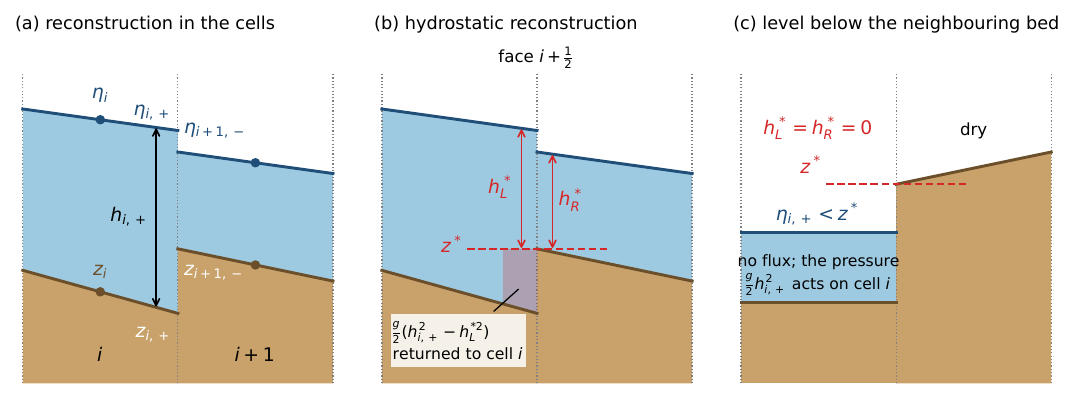}
\caption{Reconstruction in one direction (side view; water in blue, bed in brown). (a) The water level and the depth
are reconstructed linearly in each cell, and the bed follows as their difference and is therefore discontinuous
at the face. (b) At the face, the bed is taken as the higher value $z^*$ and both depths are cut off at this
level; the pressure on the part of the left state below $z^*$ (shaded) is not included in the face flux and is
returned to cell $i$. (c) If the water level of a cell lies below the bed of its neighbor, its cut-off depth
vanishes; if the neighbor is dry, both cut-off depths vanish and no water crosses the face.}\label{fig:hr}
\end{figure}

\paragraph{Reconstruction in the cell.}
The reconstructed variables are the water level $\eta$, the depth $h$ and the two components of the velocity $\mathbf{u}=(u_x,u_y)$; the bed elevation
is not reconstructed separately but follows from the first two, as in the second-order extension of Audusse et al.~\cite{audusse2004}. The velocity is reconstructed rather than the
discharge, which avoids dividing a reconstructed discharge by a small reconstructed depth. For each variable $a$,
a limited difference is
formed,
\begin{equation}
\sigma^a_i=\mathrm{minmod}\left(a_i-a_{i-1},\,a_{i+1}-a_i\right)
\end{equation}
with two modifications for shallow water: the limited difference of
the depth is further restricted to $|\sigma^h_i|\le 2h_i$, so that the depth is non-negative at both faces, and
$\sigma^{u_x}_i=\sigma^{u_y}_i=0$ if any of the three cells is shallower than $h_d$ (Section~\ref{sec:base}). The values at the two faces of the cell are
\begin{equation}
a_{i,\pm}=a_i\pm\tfrac12\sigma^a_i ,\qquad z_{i,\pm}=\eta_{i,\pm}-h_{i,\pm}.
\end{equation}
The bed elevation $z_{i,\pm}$ defined in this way varies within the cell, and its two values at a face,
$z_{i,+}$ and $z_{i+1,-}$, generally differ (Fig.~\ref{fig:hr}a). Reconstructing $\eta$ and $h$ rather than $h$ and $z$ is what keeps
still water at rest: where $\eta$ is uniform, $\sigma^\eta=0$ and the face values of the water level are equal
regardless of the bed.

\paragraph{Hydrostatic reconstruction at the face.}
At the face $i+\frac12$, the left and right states are those of the faces $(i,+)$ and $(i+1,-)$. The bed is
taken as the higher of the two values, and the depths are cut off at this level,
\begin{equation}
z^*=\max\left(z_{i,+},\,z_{i+1,-}\right),\qquad
h^*_L=\max\left(0,\ \eta_{i,+}-z^*\right),\qquad h^*_R=\max\left(0,\ \eta_{i+1,-}-z^*\right).
\end{equation}
The HLLC flux, with the components $F^h$, $F^{q_x}$ and $F^{q_y}$, is evaluated for the left state
$(h^*_L,\,h^*_L\mathbf{u}_{i,+})$ and the right state $(h^*_R,\,h^*_R\mathbf{u}_{i+1,-})$, i.e.\ with the
reconstructed velocities and the cut-off depths
(Fig.~\ref{fig:hr}b). A cell whose water level lies below the bed of its neighbor therefore presents zero depth to
the face; this is how a step or a shoreline blocks the flow (Fig.~\ref{fig:hr}c). Its water cannot pass the
face, since a state of zero depth carries no mass flux. At a shoreline the higher cell is dry, so that both
depths vanish and no water moves up onto the dry bed.

\begin{figure}[t]
\centering
\includegraphics[width=\linewidth]{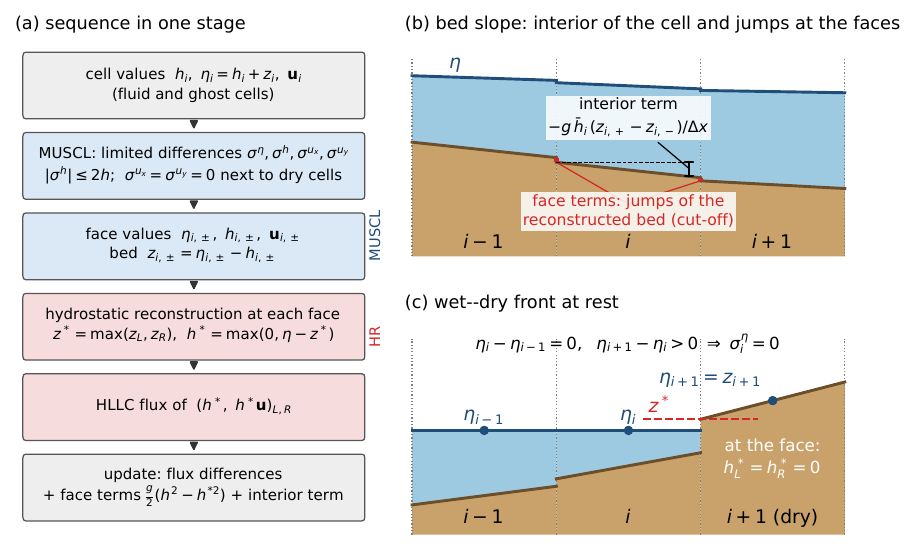}
\caption{Combination of the MUSCL reconstruction with the hydrostatic reconstruction (HR). (a) Sequence of
operations in one Runge--Kutta stage. (b) The slope of the bed is accounted for in two parts: the variation of
the reconstructed bed inside each cell (interior term) and its jumps at the faces (cut-off of the depths; the
jumps are exaggerated). (c) Wet--dry front at rest: next to a dry cell with a higher bed, the limited difference
of the water level vanishes, the water level is reconstructed as constant, and the cut-off depths at the face are zero.}
\label{fig:hr2}
\end{figure}

\paragraph{Bed slope term.}
The reconstructed bed varies linearly inside each cell and jumps at the faces. The source term
$-gh\,\partial_xz$ is split accordingly into a part at the faces and a part inside the cell
(Fig.~\ref{fig:hr2}b). The part at the faces is as follows. The momentum
flux contains the pressure $\frac12gh^2$. Since it is computed with the cut-off depth $h^*$ in place of the
reconstructed depth $h$, the pressure $\frac12g(h^2-h^{*2})$ of the water below $z^*$ is missing from it. This
water presses against the step of the bed at the face, and the step exerts the opposite force on the same water.
The missing pressure is therefore added to the momentum equation of the cell to which the depth belongs, and not
to that of its neighbor. With piecewise-constant states, the bed is constant in each cell, its entire variation
is concentrated at the faces, and this is the only part. With linear states, the bed also varies by
$z_{i,+}-z_{i,-}$ inside the cell, and the corresponding part of $-gh\,\partial_xz$ is added. It is the exact
integral over the cell for linear $h$ and $z$,
\begin{equation}
-\frac{1}{\Delta x}\int_{x_{i-1/2}}^{x_{i+1/2}} g\,h\,\partial_xz\,dx
=-\,g\,\frac{h_{i,-}+h_{i,+}}{2}\,\frac{z_{i,+}-z_{i,-}}{\Delta x}.
\end{equation}
The semi-discrete equations for cell $i$ are
\begin{align}
\frac{dh_i}{dt}&=-\frac{F^h_{i+1/2}-F^h_{i-1/2}}{\Delta x},\\
\frac{d(q_x)_i}{dt}&=-\frac{1}{\Delta x}\left[F^{q_x}_{i+1/2}+\frac g2\left(h_{i,+}^2-h^{*2}_{L,i+1/2}\right)
-F^{q_x}_{i-1/2}-\frac g2\left(h_{i,-}^2-h^{*2}_{R,i-1/2}\right)\right]\nonumber\\
&\quad+\frac g2\left(h_{i,-}+h_{i,+}\right)\frac{z_{i,-}-z_{i,+}}{\Delta x},\label{eq:hr}
\end{align}
and the transverse momentum $q_y$ receives only the flux difference. With $\sigma=0$, the last term vanishes and the first-order scheme of Audusse et al.~\cite{audusse2004} is recovered; with $\sigma\neq0$, the scheme is their second-order extension, and the last term is its centered source term. For water at rest ($\mathbf{u}=0$, $\eta$ uniform over
the wet cells), the right-hand side of (\ref{eq:hr}) vanishes, as the following four steps show. First, the water level is the
same on both sides of a face, so that both cut-off depths equal $h^*=\max(0,\eta-z^*)$; if the water level does
not reach $z^*$, both are zero. Second, the two states passed to the Riemann solver are then identical and have
no velocity. The mass flux therefore vanishes, and the momentum flux consists of the pressure alone,
$F^{q_x}=\frac12gh^{*2}$. Third, with this flux each face contributes to the bracket in (\ref{eq:hr}) the
pressure of the depth before the cut-off,
\begin{equation}
\frac g2h^{*2}_{L,i+1/2}+\frac g2\left(h_{i,+}^2-h^{*2}_{L,i+1/2}\right)=\frac g2h_{i,+}^2,\qquad
\frac g2h^{*2}_{R,i-1/2}+\frac g2\left(h_{i,-}^2-h^{*2}_{R,i-1/2}\right)=\frac g2h_{i,-}^2,
\end{equation}
and the bracket reduces to the difference of the pressures at the two ends of the cell,
$\frac g2(h_{i,+}^2-h_{i,-}^2)$. Fourth, the water level is the same at the two ends of the cell, so that the
depth increases by the amount by which the bed falls, $h_{i,+}-h_{i,-}=z_{i,-}-z_{i,+}$, and the last term of
(\ref{eq:hr}) becomes
\begin{equation}
\frac g2\left(h_{i,-}+h_{i,+}\right)\frac{h_{i,+}-h_{i,-}}{\Delta x}=\frac{1}{\Delta x}\,\frac g2\left(h_{i,+}^2-h_{i,-}^2\right).
\end{equation}
The bracket enters (\ref{eq:hr}) with the factor $-1/\Delta x$, and the two contributions cancel exactly: the
difference of the pressures at the two ends of the cell is balanced by the force of the sloping bed. The scheme is therefore well balanced: water at rest over an arbitrary bed, including a partially dry one,
remains at rest. The
immersed boundary treatment of
Section~\ref{sec:bilinear} must maintain this property, which is the reason for the dry-cell treatment of
Section~\ref{sec:drydef}: the ghost cells enter the reconstruction and the face states exactly as fluid
cells do.

\paragraph{How the two reconstructions are combined.}
The MUSCL reconstruction and the hydrostatic reconstruction are applied in sequence and serve different purposes.
The first provides second-order accurate states on the two sides of every face; the second modifies these
states so that the scheme is well balanced and the depth cannot become negative. Figure~\ref{fig:hr2}a shows the
sequence of operations in one Runge--Kutta stage: the ghost values are set first, the slopes and the face values
are computed, the depths are cut off and the HLLC flux is evaluated at every face, and the fluxes, the pressure
terms at the two faces and the interior term are assembled in (\ref{eq:hr}).

\paragraph{Wet--dry fronts.}
In a dry cell $h=0$, so that its water level $\eta=z$ is the elevation of the bed and not the level of any water. At a shoreline, this value must not enter the slope of the water level of the neighboring wet cell,
since the water level would otherwise be reconstructed as inclined and the water would start to move. The
minmod limiter prevents this. Consider water at rest and a wet cell $i$ whose neighbor $i+1$ is dry with a bed
above the water level (Fig.~\ref{fig:hr2}c). The difference to the wet neighbor, $\eta_i-\eta_{i-1}$, is zero,
and the difference to the dry neighbor, $\eta_{i+1}-\eta_i$, is positive, so that $\sigma^\eta_i=0$. The water
level of cell $i$ is therefore reconstructed as constant, both cut-off depths at the face between the two cells
are zero (Fig.~\ref{fig:hr}c), and the argument given above for water at rest applies without change.

\section{Other constructions of the ghost values}\label{sec:otherdef}

\begin{figure}[!htb]
\centering
\includegraphics[width=0.92\linewidth]{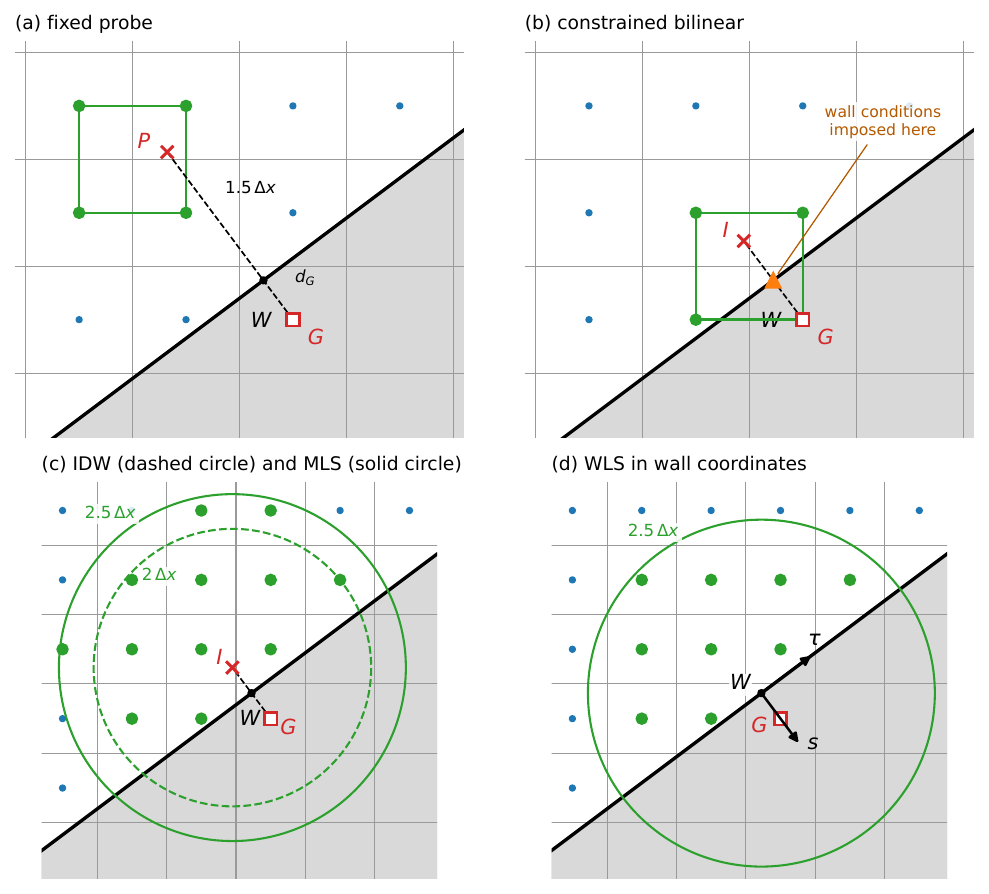}
\caption{Other constructions of the ghost values (gray: solid; blue dots: fluid cells; green dots: cells used).
(a) Fixed probe: the values are interpolated at the point $P$, at $1.5\Delta x$ from the wall. (b) Constrained
bilinear: a node around the image point that is not a fluid cell, here $G$ itself, is replaced by the wall
conditions at its projection onto the wall (triangle). (c) IDW and MLS: fluid cells within $2\Delta x$ and
$2.5\Delta x$ of the image point. (d) WLS: fluid cells within $2.5\Delta x$ of the wall point $W$, with the
coordinates $s$ and $\tau$ normal and tangential to the wall.}\label{fig:interp}
\end{figure}

This appendix describes the seven constructions of the ghost values that are compared with the baseline
(Table~\ref{tab:methods}, Fig.~\ref{fig:interp}). All of them use the same nodes as the baseline, i.e.\ the fluid
cells (those that are not dry, when the dry-cell treatment is used), and the notation of Section~\ref{sec:bilinear}: $G$ is the ghost cell, $W$ the wall point,
$I$ the image point and $d_G$ the distance of $G$ from the wall. Within the proposed scheme, the dry-cell
treatment of Section~\ref{sec:drydef} is applied to their weights in the same way.

\paragraph{Fixed probe.}
The interpolation point $P$ lies on the normal through $G$ at the fixed distance $d_P=1.5\Delta x$ from the wall,
whereas the image point lies at the distance $d_G$. The values at $P$ are interpolated as in the baseline. The
water level and the tangential discharge are copied to the ghost cell, and the normal discharge is extrapolated
along the straight line that passes through zero at the wall,
\begin{equation}
\eta_G=\eta_P,\qquad
\mathbf{q}_G=\left[\mathsf{I}-\left(1+\frac{d_G}{d_P}\right)(\mathbf{n}\otimes\mathbf{n})\right]\mathbf{q}_P .
\end{equation}
With $d_P=d_G$ this is the baseline.

\paragraph{Constrained bilinear.}
The values at the image point are obtained from a bilinear function $a+bx+cy+dxy$ on the square formed by the
four nodes around $I$. Its four coefficients are determined by four conditions, one from each node. At a fluid
node, the function equals the value of the cell. A node that is not a fluid cell is not dropped as in the
baseline but replaced by a wall condition, imposed at the projection of the node onto the wall (triangle in
Fig.~\ref{fig:interp}b): $\partial\eta/\partial n=0$ for the water level, and $q_n=0$ and
$\partial q_t/\partial n=0$ for the discharge \cite{mittal2008,rinco2012}. Since $q_n$ and $q_t$ combine the two
components of the discharge, the two components cannot be determined separately: a $4\times4$ system is solved
for the water level and an $8\times8$ system for the discharge. The function is evaluated at $I$ and mirrored as
in the baseline. Where the system is singular, the baseline is used.

\paragraph{IDW.}
The value at $I$ is the mean of the values of the fluid cells within $2\Delta x$ of $I$, weighted by the inverse square of
their distance from $I$ \cite{shepard1968,gao2007}; it is mirrored as in the baseline.

\paragraph{MLS1 and MLS2.}
A polynomial of degree one (three coefficients) or two (six coefficients) is fitted by weighted least squares to
the fluid cells within $R=2.5\Delta x$ of $I$ \cite{lancaster1981,vanella2009}. A cell at the distance $r$ from
$I$ has the weight $(1-r/R)^4(1+4r/R)$. The polynomial is evaluated at $I$ and mirrored as in the baseline. If
there are fewer cells than coefficients, or if the system is singular, the degree is lowered.

\paragraph{WLS1 and WLS2.}
No image point is used. Instead, a function that satisfies the wall conditions for any choice of its coefficients
is fitted to the fluid cells near the wall and evaluated at the ghost cell itself. The coordinates are $s$ along
$\mathbf{n}$ and $\tau$ along $\mathbf{t}$, with origin at $W$, so that the wall is at $s=0$ and the ghost cell
at $s=d_G$. For the water level and the tangential discharge the function is $a+b\tau$ (degree one) or
$a+b\tau+c\tau^2+ds^2$ (degree two); it has no term linear in $s$, so that its normal gradient vanishes at the
wall. For the normal discharge it is $as$ (degree one) or $as+bs\tau+cs^2$ (degree two), which vanishes at the
wall. The coefficients are determined by weighted least squares from the fluid cells within $2.5\Delta x$ of $W$,
with the weight given above.


\begin{thebibliography}{99}
\bibitem{audusse2004} E.~Audusse, F.~Bouchut, M.-O.~Bristeau, R.~Klein, B.~Perthame, A fast and stable well-balanced scheme with hydrostatic reconstruction for shallow water flows, SIAM J. Sci. Comput. 25~(6) (2004) 2050--2065. doi:10.1137/S1064827503431090
\bibitem{berger2021} M.~Berger, A.~Giuliani, A state redistribution algorithm for finite volume schemes on cut cell meshes, J. Comput. Phys. 428 (2021) 109820. doi:10.1016/j.jcp.2020.109820
\bibitem{canestrelli2016} A.~Canestrelli, A.~Spruyt, B.~Jagers, R.~Slingerland, M.~Borsboom, A mass-conservative staggered immersed boundary model for solving the shallow water equations on complex geometries, Int. J. Numer. Methods Fluids 81~(3) (2016) 151--177. doi:10.1002/fld.4180
\bibitem{causon2000} D.~M.~Causon, D.~M.~Ingram, C.~G.~Mingham, G.~Yang, R.~V.~Pearson, Calculation of shallow water flows using a Cartesian cut cell approach, Adv. Water Resour. 23~(5) (2000) 545--562. doi:10.1016/S0309-1708(99)00036-6
\bibitem{gottlieb2001} S.~Gottlieb, C.-W.~Shu, E.~Tadmor, Strong stability-preserving high-order time discretization methods, SIAM Rev. 43~(1) (2001) 89--112. doi:10.1137/S003614450036757X
\bibitem{mittal2005} R.~Mittal, G.~Iaccarino, Immersed boundary methods, Annu. Rev. Fluid Mech. 37 (2005) 239--261. doi:10.1146/annurev.fluid.37.061903.175743
\bibitem{mittal2008} R.~Mittal, H.~Dong, M.~Bozkurttas, F.~M.~Najjar, A.~Vargas, A.~von~Loebbecke, A versatile sharp interface immersed boundary method for incompressible flows with complex boundaries, J. Comput. Phys. 227~(10) (2008) 4825--4852. doi:10.1016/j.jcp.2008.01.028
\bibitem{ouro2016} P.~Ouro, L.~Cea, L.~Ram\'{\i}rez, X.~Nogueira, An immersed boundary method for unstructured meshes in depth averaged shallow water models, Int. J. Numer. Methods Fluids 81~(11) (2016) 672--688. doi:10.1002/fld.4201
\bibitem{rinco2012} S.~Rinco, Metodi immersed boundary ai volumi finiti centrali per il sistema shallow water, Tesi di laurea magistrale (supervisor: E.~Miglio), Politecnico di Milano, 2012. http://hdl.handle.net/10589/51682
\bibitem{ryoo2021} C.~Ryoo, K.~Mandli, Barrier simulations with shallow water equations using the state redistribution method, arXiv:2112.09696 [math.NA] (2021).
\bibitem{seo2014} I.~W.~Seo, Y.~D.~Kim, C.~G.~Song, Validation of depth-averaged flow model using flat-bottomed benchmark problems, Sci. World J. 2014 (2014) 197539. doi:10.1155/2014/197539
\bibitem{shu1988} C.-W.~Shu, S.~Osher, Efficient implementation of essentially non-oscillatory shock-capturing schemes, J. Comput. Phys. 77~(2) (1988) 439--471. doi:10.1016/0021-9991(88)90177-5
\bibitem{toro1994} E.~F.~Toro, M.~Spruce, W.~Speares, Restoration of the contact surface in the HLL-Riemann solver, Shock Waves 4~(1) (1994) 25--34. doi:10.1007/BF01414629
\bibitem{toro2001} E.~F.~Toro, Shock-Capturing Methods for Free-Surface Shallow Flows, John Wiley \& Sons, 2001. ISBN 978-0-471-98766-6.
\bibitem{tseng2003} Y.-H.~Tseng, J.~H.~Ferziger, A ghost-cell immersed boundary method for flow in complex geometry, J. Comput. Phys. 192~(2) (2003) 593--623. doi:10.1016/j.jcp.2003.07.024
\bibitem{vanleer1979} B.~van~Leer, Towards the ultimate conservative difference scheme. V. A second-order sequel to Godunov's method, J. Comput. Phys. 32~(1) (1979) 101--136. doi:10.1016/0021-9991(79)90145-1
\bibitem{chern1987} I.-L.~Chern, P.~Colella, A conservative front tracking method for hyperbolic conservation laws, Technical Report UCRL-97200, Lawrence Livermore National Laboratory, 1987.
\bibitem{fadlun2000} E.~A.~Fadlun, R.~Verzicco, P.~Orlandi, J.~Mohd-Yusof, Combined immersed-boundary finite-difference methods for three-dimensional complex flow simulations, J. Comput. Phys. 161~(1) (2000) 35--60. doi:10.1006/jcph.2000.6484
\bibitem{gao2007} T.~Gao, Y.-H.~Tseng, X.-Y.~Lu, An improved hybrid Cartesian/immersed boundary method for fluid--solid flows, Int. J. Numer. Methods Fluids 55~(12) (2007) 1189--1211. doi:10.1002/fld.1522
\bibitem{lai2011} W.~Lai, A.~A.~Khan, A discontinuous Galerkin method for two-dimensional shock wave modeling, Model. Simul. Eng. 2011 (2011) 782832. doi:10.1155/2011/782832
\bibitem{lancaster1981} P.~Lancaster, K.~Salkauskas, Surfaces generated by moving least squares methods, Math. Comput. 37~(155) (1981) 141--158. doi:10.1090/S0025-5718-1981-0616367-1
\bibitem{liang2007} Q.~Liang, J.~Zang, A.~G.~L.~Borthwick, P.~H.~Taylor, Shallow flow simulation on dynamically adaptive cut cell quadtree grids, Int. J. Numer. Methods Fluids 53~(12) (2007) 1777--1799. doi:10.1002/fld.1363
\bibitem{liang2008} Q.~Liang, A.~G.~L.~Borthwick, Simple treatment of non-aligned boundaries in a Cartesian grid shallow flow model, Int. J. Numer. Methods Fluids 56~(11) (2008) 2091--2110. doi:10.1002/fld.1615
\bibitem{pember1995} R.~B.~Pember, J.~B.~Bell, P.~Colella, W.~Y.~Crutchfield, M.~L.~Welcome, An adaptive Cartesian grid method for unsteady compressible flow in irregular regions, J. Comput. Phys. 120 (1995) 278--304. doi:10.1006/jcph.1995.1165
\bibitem{seo2011} J.~H.~Seo, R.~Mittal, A sharp-interface immersed boundary method with improved mass conservation and reduced spurious pressure oscillations, J. Comput. Phys. 230~(19) (2011) 7347--7363. doi:10.1016/j.jcp.2011.06.003
\bibitem{shepard1968} D.~Shepard, A two-dimensional interpolation function for irregularly-spaced data, in: Proceedings of the 1968 23rd ACM National Conference, 1968, pp.~517--524. doi:10.1145/800186.810616
\bibitem{vanella2009} M.~Vanella, E.~Balaras, A moving-least-squares reconstruction for embedded-boundary formulations, J. Comput. Phys. 228~(18) (2009) 6617--6628. doi:10.1016/j.jcp.2009.06.003
\bibitem{an2012} H.~An, S.~Yu, Well-balanced shallow water flow simulation on quadtree cut cell grids, Adv. Water Resour. 39 (2012) 60--70. doi:10.1016/j.advwatres.2012.01.003
\bibitem{causon2001} D.~M.~Causon, D.~M.~Ingram, C.~G.~Mingham, A Cartesian cut cell method for shallow water flows with moving boundaries, Adv. Water Resour. 24~(8) (2001) 899--911. doi:10.1016/S0309-1708(01)00010-0
\bibitem{chung2011} H.-R.~Chung, T.-Y.~Hsieh, J.-C.~Yang, Two dimensional shallow-water flow model with immersed boundary method, Comput. Fluids 51~(1) (2011) 145--156. doi:10.1016/j.compfluid.2011.08.009
\bibitem{derosis2014} A.~De~Rosis, A lattice Boltzmann-finite element model for two-dimensional fluid--structure interaction problems involving shallow waters, Adv. Water Resour. 65 (2014) 18--24. doi:10.1016/j.advwatres.2014.01.003
\bibitem{kevlahan2015} N.~K.-R.~Kevlahan, T.~Dubos, M.~Aechtner, Adaptive wavelet simulation of global ocean dynamics using a new Brinkman volume penalization, Geosci. Model Dev. 8~(12) (2015) 3891--3909. doi:10.5194/gmd-8-3891-2015
\bibitem{kim2011} H.-J.~Kim, Y.-S.~Cho, Numerical model for flood routing with a Cartesian cut-cell domain, J. Hydraul. Res. 49~(2) (2011) 205--212. doi:10.1080/00221686.2010.547037
\bibitem{kwon2021} H.~J.~Kwon, H.~Lee, S.-M.~Chang, High-order WENO schemes with an immersed boundary method for shallow water equations on the tsunami mitigation with configurations of cylinder array, KSCE J. Civ. Eng. 25~(1) (2021) 1--11. doi:10.1007/s12205-020-0535-y
\bibitem{li2021} J.~Li, K.~T.~Mandli, An $h$-box method for shallow water equations including barriers, SIAM J. Sci. Comput. 43~(2) (2021) B431--B454. doi:10.1137/19M128363X
\bibitem{liang2008b} Q.~Liang, Simulation of shallow flows in nonuniform open channels, J. Fluids Eng. 130~(1) (2008) 011205. doi:10.1115/1.2829593
\bibitem{ning2008} D.~Z.~Ning, J.~Zang, Q.~Liang, P.~H.~Taylor, A.~G.~L.~Borthwick, Boussinesq cut-cell model for non-linear wave interaction with coastal structures, Int. J. Numer. Methods Fluids 57~(10) (2008) 1459--1483. doi:10.1002/fld.1656
\bibitem{oler2016} A.~Oler, N.~Zhang, S.~R.~Brandt, Q.~Chen, Implementation of an infinite-height levee in CaFunwave using an immersed-boundary method, J. Fluids Eng. 138~(11) (2016) 111103. doi:10.1115/1.4033490
\bibitem{prestininzi2016} P.~Prestininzi, V.~Lombardi, M.~La~Rocca, Curved boundaries in multi-layer shallow water lattice Boltzmann methods: bounce back versus immersed boundary, J. Comput. Sci. 16 (2016) 16--28. doi:10.1016/j.jocs.2016.03.001
\bibitem{reckinger2012} S.~M.~Reckinger, O.~V.~Vasilyev, B.~Fox-Kemper, Adaptive volume penalization for ocean modeling, Ocean Dyn. 62~(8) (2012) 1201--1215. doi:10.1007/s10236-012-0555-3
\bibitem{rosatti2005} G.~Rosatti, D.~Cesari, L.~Bonaventura, Semi-implicit, semi-Lagrangian modelling for environmental problems on staggered Cartesian grids with cut cells, J. Comput. Phys. 204~(1) (2005) 353--377. doi:10.1016/j.jcp.2004.10.013
\bibitem{ryoo2021b} C.~Ryoo, K.~Mandli, Barrier simulations and experimental calculations using cell merging method, arXiv:2112.09566 (2021).
\bibitem{song2018} T.~Song, A.~Main, G.~Scovazzi, M.~Ricchiuto, The shifted boundary method for hyperbolic systems: embedded domain computations of linear waves and shallow water flows, J. Comput. Phys. 369 (2018) 45--79. doi:10.1016/j.jcp.2018.04.052
\bibitem{taylor2026} C.~G.~Taylor, J.~Chan, An entropy stable high-order discontinuous Galerkin method on cut meshes, J. Comput. Phys. 548 (2026) 114551. doi:10.1016/j.jcp.2025.114551
\bibitem{zhou2004} J.~G.~Zhou, D.~M.~Causon, C.~G.~Mingham, D.~M.~Ingram, Numerical prediction of dam-break flows in general geometries with complex bed topography, J. Hydraul. Eng. 130~(4) (2004) 332--340. doi:10.1061/(ASCE)0733-9429(2004)130:4(332)
\bibitem{brufau2000} P.~Brufau, P.~Garc\'{\i}a-Navarro, Two-dimensional dam break flow simulation, Int. J. Numer. Methods Fluids 33~(1) (2000) 35--57. doi:10.1002/(SICI)1097-0363(20000515)33:1$<$35::AID-FLD999$>$3.0.CO;2-D
\end{thebibliography}
\end{document}